\documentclass{elsart}
\usepackage{natbib}
\bibpunct{[}{]}{,}{n}{}{;}
\usepackage{color}
\usepackage{graphicx}
\usepackage{amssymb}
\usepackage{amsmath}
\usepackage{algorithmic}
\usepackage{epsf}
\usepackage{tabularx}

\newcommand{\eg}{{\em e.g., }}           
\newcommand{\ie}{{\em i.e., }}           

\def\be{\begin{equation}}
\def\ee{\end{equation}}
\def\ba{\begin{array}}
\def\ea{\end{array}}
\def\d{\mathrm{d}}

\newcommand{\revision}[1]{\textcolor{black}{#1}}

\begin{document}

\begin{frontmatter}

\title{Three dimensional image correlation from X-Ray computed
tomography of solid foam}

\author{St\'{e}phane Roux},
\ead{stephane.roux@lmt.ens-cachan.fr}
\author{Fran\c{c}ois Hild},

\address{Laboratoire de M\'ecanique et Technologie (LMT-Cachan)\\
                 Ecole Normale Sup\'erieure de Cachan / CNRS-UMR 8535 /
                 Universit\'e Paris 6\\ 61 Avenue du Pr\'esident Wilson,
                 F-94235 Cachan Cedex, France.}

\author{Philippe Viot}
\address{LAMEFIP, ENSAM de Bordeaux,\\
Esplanade des Arts et M\'{e}tiers, 33405 Talence Cedex, France. }
\and
\author{Dominique Bernard}
\address{Institut de Chimie de la Mati\`{e}re Condens\'{e}e de Bordeaux
ICMCB,\\
CNRS/Universit\'{e} Bordeaux I,\\ 87, Av. Dr A.Schweitzer, 33608
PESSAC Cedex, France.}

\begin{abstract}
A new methodology is proposed to estimate 3D displacement fields
from pairs of images obtained from X-Ray Computed Micro
Tomography (XCMT). Contrary to local approaches, a global
approach is followed herein that evaluates {\em continuous}
displacement fields. Although any displacement basis could be
considered, the procedure is specialized to finite element shape
functions. The method is illustrated with the analysis of a
compression test on a polypropylene solid foam (independently
studied in a companion paper). A good stability of the measured
displacement field is obtained for cubic element sizes ranging
from $16$ voxels to $6$ voxels.
\end{abstract}


\end{frontmatter}

\section{Introduction}


X-Ray Computed Micro Tomography (XCMT) is a very precious tool to
have access to the details of the full microstructure of a
material in a non destructive fashion. By reconstruction from 2D
pictures, it allows for a 3D visualization of the different
phases of a material~\cite{1979,1971}. By analyzing these 3D
reconstructed pictures, one has access, for instance, to
microstructural changes during solidification of
alloys~\cite{1967}, to microstructural details during sintering
of steel powders~\cite{1970,bernard}, to damage mechanisms in the
bulk of particulate composites~\cite{1973,1972} or to the
structure of cellular materials (either metallic or polymeric
foams) and strains~\cite{1975,1980}. These actual microstructures
may be further processed using e.g. finite element
tools~\cite{1985} for additional exploitations.

Composite materials are of particular interest for the
above-mentioned type of studies since the combination of
different phases and materials calls for modeling tools that
account for strain and stress distributions along and close to
interfaces and/or interphases. For instance, damage mechanisms
(\eg interface debonding, fiber breakage) were analyzed in
fiber-reinforced composites~\cite{1982}. The load transfer
between fibers and matrices in the presence of cracks was also
investigated, and mechanical properties of interfaces were then
identified~\cite{1984}. The strain field was obtained by X-ray
diffraction techniques.

The behavior of composite sandwich panels is often studied since
these structures are widely used in transportation industries.
Under extreme conditions (\eg accidents or crashes), these
structures are designed to absorb the impact energy to ensure
passengers safety. Therefore, the dynamic response of sandwich
structures has to be identified at high strain rates. The
essential role of the core on the composite response has been
demonstrated~\cite{2002}, and the macroscopic mechanical behavior
of different types of core (wood, metallic or polymeric foams)
has been identified~\cite{2003}. The behavior of cellular
materials generally includes three steps in static or dynamic
compression, namely, an elastic response followed by an important
deformation of the material with quasi-constant stress due to
strain localization, and densification. The classical models,
which are used to predict and improve the behavior of these
structures under impact, describe the macroscopic response but do
not account for strain localization and hence are unable to
relate the behavior to the local microstructure~\cite{2004}. The
cellular material under investigation is a multi-scale foam made
out of millimetric porous beads (at the meso scale) which
themselves consist of closed micro cells. To improve this foam
used as core in sandwich panel, it is therefore essential to
propose macroscopic constitutive equations that take into account
the physical phenomena observed at all scales (macro-, meso- and
micro-scales) of the foam structure. A first point of this study
is to reveal strain localization at the bead scale from 3D images
obtained by XCMT. Second, strain localization within the bead
structure is investigated at the microscopic scale. Last, the
measurement of the 3D strain field opens the way for a realistic
description incorporating microstructural information, and hence
in the long term to optimize this microstructure to meet
mechanical requirements for best safety performances.

Cellular materials are one of the constituents of shock absorbing
composite~\cite{2003}. In a companion paper~\cite{1974}, the
potential of a direct and exhaustive characterization of the
microstructure of a multiscale foam after impact loadings is
discussed. Morphology of bead wall was shown to reveal buckling
phenomena. Moreover, the use of mathematical morphology tools
enables for the extraction individual beads, the estimation of
their densities and of the bead density distribution compared to
volumetric strains.  However, for further information on the
strain at a finer scale require a different approach.

In the case of structural composites, 2D Digital Image
Correlation (2D-DIC) techniques were used (with images obtained
by using a CCD camera) to measure strain fields, to study the
influence of the fiber orientation on the macroscopic behavior or
to reveal strain and damage localization at the finest
scales~\cite{1991,1336,1989,1990}. Few studies have reported
detailed investigations of kinematic analyses using 3D pictures.
As a complementary analysis, X-ray diffraction may be used to
evaluate elastic strains~\cite{1984}. However, the reconstructed
pictures themselves may be used to evaluate displacements. A
first technique consists in tracking markers (\eg particles).
Strain uncertainties of the order of $10^{-2}$ have been
reported~\cite{1983}. When image processing techniques based upon
correlation algorithms are used, strain uncertainties of the
order of $10^{-3}$ can be achieved~\cite{1912,1796}. A major
field of application concerns biomechanical
studies~\cite{1911,1968}. Other studies were devoted to
micromechanical analyses of strain fields in heterogeneous
materials~\cite{1796,1970}. Cellular materials have also been
studied~\cite{1975,1980}, and 2D strain measurements were
performed~\cite{1977}.

The final objective of the present study is to improve mechanical
models of sandwich structures (with composite face-sheets and
foam core) under dynamic loading. In this first step, 3D-DIC and
tomographic techniques are applied to better identify the
behavior of the foam core and analyze strain localization
appearing in cellular materials during impact. This step is
achieved by resorting to correlation techniques. Digital image
correlation is a technique that consists in measuring
displacement fields based on image pairs of the same specimen at
different stages of loading~\cite{1323}. The displacement field
is computed so that the image of the strained sample is matched
to the reference image when pixel locations are corrected by the
displacement field. Very often, surface images are taken so that
only the in-plane displacement and strain components can be
identified when using one camera (2D-DIC) or 3D displacements and
2D strains of the surface by using stereo-correlation (2.5-DIC).
This technique has revealed to be extremely powerful in solid
mechanics for different reasons. It is easy to perform with
digital cameras that have very good performances and are low
cost, it is a non-intrusive and very tolerant technique that does
not require heavy specimen preparations, and last it provides a
very rich information (\ie full-fields). Moreover, algorithms
have progressed in the recent years, providing today both robust
and very accurate tools (displacement uncertainties are of order
$10^{-2}$~pixel or even below). Last, it allows one to bridge the
gap between experiments and simulations by using identification
techniques that explicitly use full-field data~\cite{1761} or for
validation purposes of numerical codes~\cite{1944}.

As in 2D applications, the most commonly used correlation
algorithms consist in matching {\em locally} small zones of
interest in a sequence of pictures to determine local displacement
components~\cite{1919}. The same type of hypotheses are made in
three-dimensional algorithms~\cite{1912,1969,1796,1906}. This
study reports on a different algorithm, which has been shown to be
more performing in two dimensions when compared to FFT-based
algorithms~\cite{1847,1818}, but had never been extended to three
dimensions. The basic approach is close in spirit to the two
dimensional case. However, specific procedures aim at saving
memory storage, which appears today to be the limiting factor,
when using conventional PCs. Since 3D images are the result of a
computed reconstruction, the image texture, and more generally its
quality cannot directly be compared to two-dimensional optical
images. Moreover, space dimensionality has an impact on the
expected performances.  It is therefore important to validate the
performance of the algorithm, one of the motivations of the
present paper.

The paper is organized as follows. In Section~\ref{se-exp}, the
way the pictures were obtained through scanning in-situ a
compressed foam is discussed. The 3D correlation algorithm is
introduced in Section~\ref{se-C8DIC} to analyze the obtained
pictures. A ``global'' procedure is implemented in which the
displacement field is decomposed over a basis made of shape
functions of 8-node finite elements. The performance of the
algorithm is evaluated in Section~\ref{se-perfo} by artificially
translating the {\em actual} reference picture of
Section~\ref{se-exp}. In Section~\ref{se-results}, first results
are discussed when considering a deformed picture. In particular,
the heterogeneity of the displacement and strain fields is
analyzed.

\section{Description of the experimental test}
\label{se-exp}

\subsection{Material}

The material under investigation is a polypropylene foam. It is a
multi-scale material made out of millimetric beads that are
agglomerated. During processing, the expanded plastic foam beads
are injected into a mold where individual beads are fused
together under steam heat and pressure to form a medium without
inter-bead porosity. Moreover, the beads themselves are cellular
structures with closed cells of size ranging from 10 to
60~$\mu$m, with a rather homogeneous microstructure, apart from a
much denser skin of about 10~$\mu$m in thickness, deprived of
large bubbles. As the material is formed, the skin of the beads
come together and merge to form a large scale cellular structure.
The foam structure is therefore made of beads (mesoscale) and
cells (microscale).

\begin{figure}
 \centerline{a)\epsfxsize0.7\textwidth \epsffile{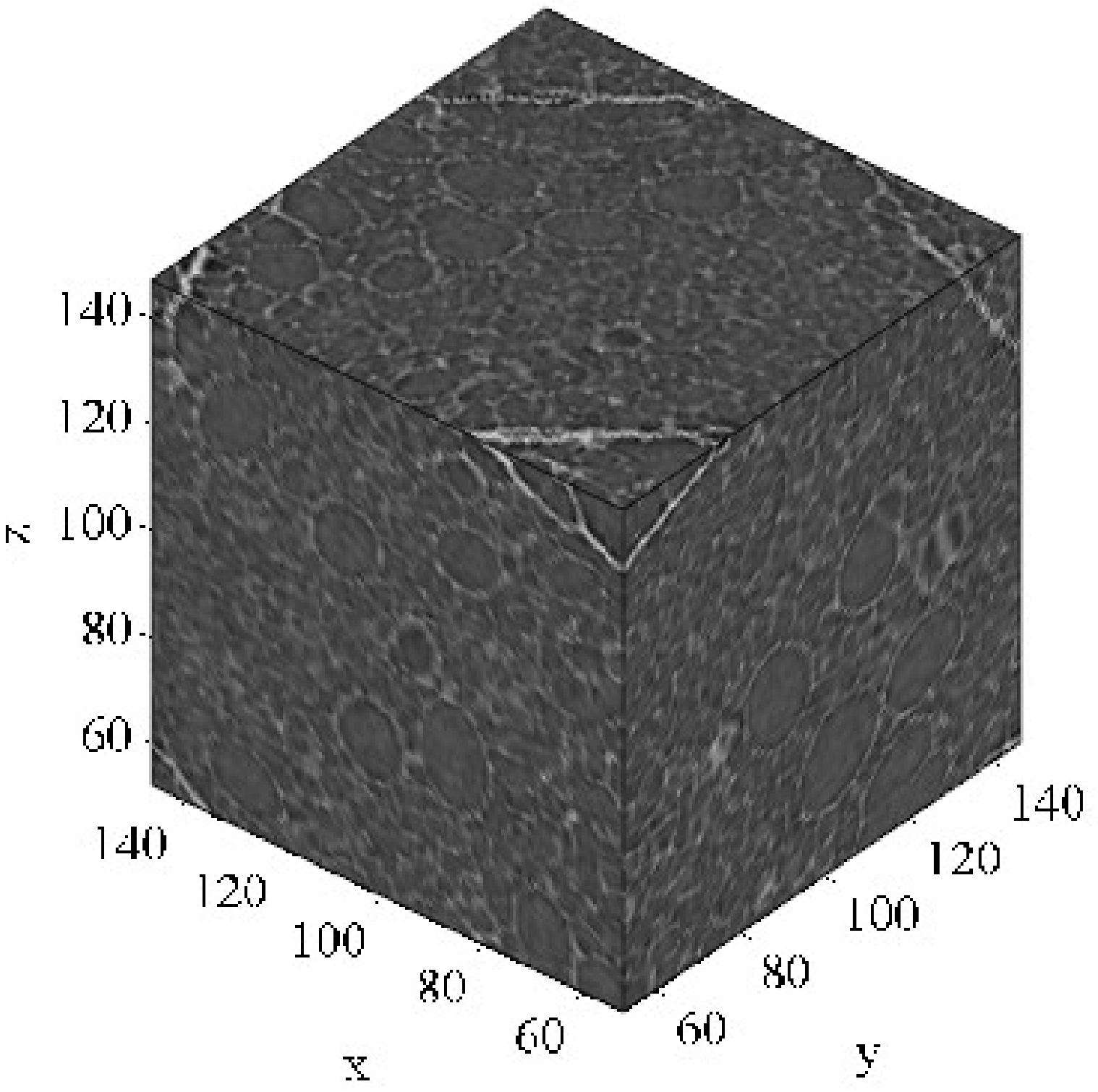}}
 \centerline{b)\epsfxsize0.7\textwidth \epsffile{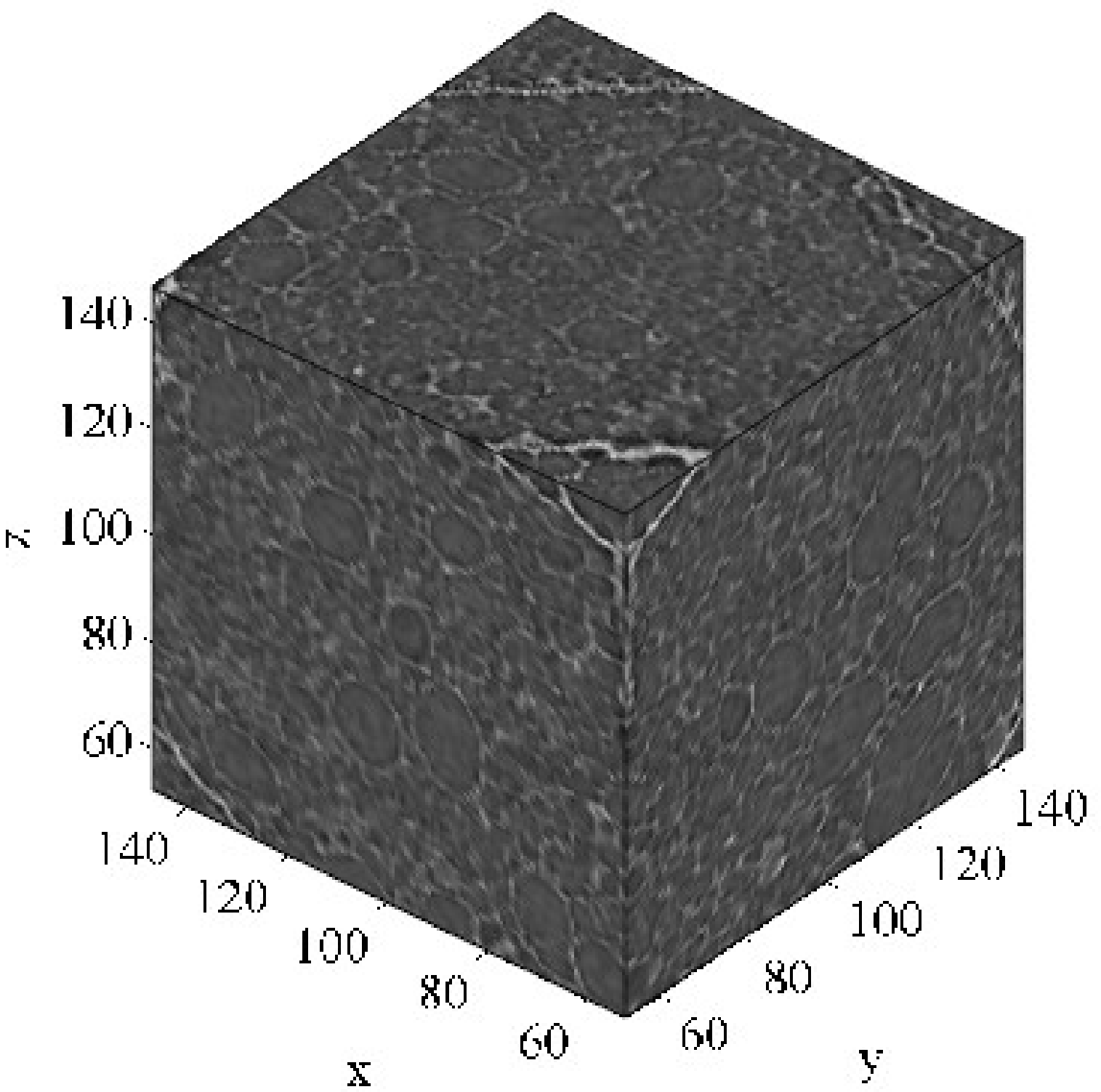}}
  \caption{View of the $96$~voxel element volume
  analyzed in this study (a: reference, b: deformed configurations).
  A 14-bit deep grey level digitization is used.}
  \label{fig:specimen}
\end{figure}

Microtomographic images presented in this paper have been
obtained on the BM05 beam line at the European Synchrotron
Radiation Facility (ESRF) in Grenoble (France). The acquired
projections are $2048~\times~2048$~pixel radiographs, with a
pixel corresponding to 4.91~$\mu$m. The sample of polypropylene
foam is 10 mm in diameter, and the energy of the beam was set to
16~keV. The use of microtomographic techniques needs a specific
methodology to measure the state of the foam structure during
dynamic loading. The adopted experimental approach consists in
carrying out several interrupted impact tests on a same sample
using a drop tower, and acquiring a ``picture'' in between two
impact tests~\cite{1975}. A tomographic measurement of the sample
is taken before the first impact and is used as a reference.
During each compression, the deformation amplitude is limited to
fixed values, namely 1~mm for the first impact, and 2~mm for the
following ones. The sample is maintained compressed and replaced
in the microtomographic setup for another acquisition. These
operations (impact and X-ray scan) are repeated until
densification of the foam. The cellular material deformation can
then be evaluated from 3D reconstructions at each stage of the
dynamic test (Figure~\ref{fig:specimen}).

\subsection{Connection with other related studies}

Most studies focused on the influence of the microstructure (\eg
cell shape, size distribution) on the macroscopic behavior of the
foam. The influence of the intermediate scale is seldom
investigated. However, considering the thickness of bead walls
and their relatively high density, it appears necessary to take
into account the bead wall rigidity, and further its buckling
combined to the cell deformation in order to accurately describe
the macroscopic behavior of foams. The microstructure of the
material is suitable for the present purpose since it has density
contrasts at different scales (from the bubble size to the bead
size) that are used here as random ``markers'' for the
correlation algorithm.

Strain fields of the bead (first, along a vertical
plane~\cite{1980} and then in the bulk of the whole
sample~\cite{1974}) have highlighted the physical phenomena of
polymeric foam damage under dynamic compression. There are three
stages in the deformation of cellular material in compression.
First, after a transient elastic phase, stronger bead deformation
is located near the compression device surface. In a second
phase, observed after the second and third impacts, strains are
localized on layers generally perpendicular to the compression
direction. Last, densification occurs.

The interrupted impact methodology coupled with micro-tomographic
analysis have shown a relationship between bead density and
strain level. However, the results show also that the strain
heterogeneity is not simply correlated with the density field,
namely, the densest beads are not necessarily the least deformed
zones. The most porous beads generate randomly located weakness
zones in the cellular material that may initiate a local collapse
if the neighborhood is favorable, and finally produce a densified
layer traversing the entire specimen. It is thus the combined
effect of the structure morphology and of the heterogeneous
density field that creates a specific strain field for a dynamic
compression. There is therefore a need for a complete 3D strain
fields to better understand the correlation between the material
texture and the way the material deforms.

\section{C8 Digital Image Correlation (C8-DIC)}
\label{se-C8DIC}

Digital Image Correlation is based on the principle of optical
flow conservation~\cite{1399,1400}, namely, the texture of the
medium is assumed to be passively advected by the displacement
field.  If the reference (respectively deformed) image is
represented by a three dimensional gray level-valued field,
$f(\mathbf{x})$, (respectively $g(\mathbf {x})$), optical flow
conservation requires that
 \be
 g(\mathbf{x})=f(\mathbf{x}+\mathbf{U}(\mathbf{x}))
 \ee
Let us underline that this hypothesis is only an approximation
that may be in default in particular for large strains, where the
gray value is an integral measure over a voxel size. Moreover, in
the case of tomography, a number of artifacts affect the
reconstruction, and violate the above hypothesis. However, taking
into account a different law, although possible in principle,
would require a specific analysis of the bias in the data
acquisition technique for the full image, and in the sequel it is
assumed that the above hypothesis holds.

In the present study, the general strategy that consists in
exploiting cross-correlations between small zones of
interest~\cite{1912,1969,1796,1906} is not used.  Rather, an
extension to three dimensions of a numerical scheme, which is
presented in details for two dimensions in Ref.~\cite{1847}) is
proposed. It is based on a {\em continuous} and global field as
commonly practiced in finite element simulations~\cite{223}. It
will therefore allow, for example, for {\em direct and unbiased}
comparisons with finite element simulations when the same
kinematic hypotheses are made during the measurement and the
simulation stages. The basic principles are recalled here, as well
as the specific modifications that have been considered for the
sake of reducing memory and/or time usage.

The optical flow conservation is considered under the weak form
of minimizing the following objective functional ${\mathcal T}$
  \be
  {\mathcal T}(\mathbf{U})=
  \int\!\!\!\!\int\!\!\!\!\int_D
  [g(\mathbf{x})-f(\mathbf{x}+\mathbf{U}(\mathbf{x}))]^2
  \d \mathbf{x}
  \ee
This functional is strongly non-linear and generally displays
numerous secondary minima.  For computational efficiency, one
resorts to a linearized function $\widetilde{\mathcal T}$
  \be\label{eq:obj_lin}
  \widetilde{\mathcal T}(\mathbf{U})=
  \int\!\!\!\!\int\!\!\!\!\int_D
  [g(\mathbf{x})-f(\mathbf{x})-\mathbf{\nabla}
  f(\mathbf{x}).\mathbf{U}(\mathbf{x})]^2 \d \mathbf{x}
  \ee
The way secondary spurious minima are dealt with is two-fold:
\begin{itemize}
\item first, the displacement field is searched in a restricted
space of functions,
\item second, the texture of the images $f$ and $g$ is severely
filtered in such a way that the Taylor expansion provides a
consistent expression of the texture variation at the scale of
the computed displacement.
\end{itemize}

Once a first determination of the displacement has been computed,
the deformed image is corrected by this first determination in
order to match the reference one. Because of the filtering, and
the restricted nature of the displacement field that is
considered, one expects that the displacement field be only a
gross approximation of the actual one.  However, once the first
correction has been performed, the remaining displacement field is
likely to be of much smaller amplitude.  Consequently, one
re-iterates the same procedure, with a less filtered image, and an
enriched displacement function basis.  This defines one step of an
algorithm that may be carried out up to the stage where the
unfiltered image is used. In spirit, it is close to multi-grid
algorithms used in 2D digital image
correlation~\cite{1326,1847,1976} that allow for the measurement
of large deformations.

This general procedure may be applied to a variety of different
fields and filters. In the sequel, the considered displacement
basis is specialized to be finite element functions over 8-node
cubic elements and piecewise tri-linear functions (C8P1 elements).
It follows that the algorithm is referred to as C8-DIC. Let us
introduce the decomposition
  \be
  \mathbf{U}(\mathbf{x})=\sum_i a_i {\mathbf{N}}_i(\mathbf{x})
  \ee
The minimization of the quadratic functional~(\ref{eq:obj_lin})
leads to the following linear system
  \be
  M_{ij}a_j=b_i
  \ee
where
  \be\ba{ll}
  M_{ij}&\displaystyle=\int\!\!\!\!\int\!\!\!\!\int
  [\mathbf{\nabla} f(\mathbf{x}).{\mathbf{N}}_i(\mathbf{x})]
  [\mathbf{\nabla} f(\mathbf{x}).{\mathbf{N}}_j(\mathbf{x})]
  \d \mathbf{x}\\
  b_i&\displaystyle=\int\!\!\!\!\int\!\!\!\!\int
  [\mathbf{\nabla} f(\mathbf{x}).{\mathbf{N}}_i(\mathbf{x})]
  [g(\mathbf{x})-f(\mathbf{x})] \d \mathbf{x}\\
  \ea\ee
Thus a single parameter characterizes the fineness of the
displacement basis, namely the size $\ell$ (in voxels) of the
element used.  The latter is comparable to the so-called Zone Of
Interest (ZOI) size of classical DIC codes.

In so-doing, the displacement field is continuous, and this is an
upgrade compared to the procedure used in the early days of 2D
digital image correlation~\cite{1175} where the displacement
field was piecewise constant. It was shown in two dimensions that
this continuity allowed one to use much smaller zones of interest
(typically 8 to 16-pixel as compared to 32-pixel
ZOIs~\cite{1847}). In terms of filters, the most elementary one
is used as for the two-dimensional procedure~\cite{1326}, namely,
a coarse-graining procedure in which ``macro-voxels'' are defined
to be the sum of $2 \times 2 \times 2$ elementary voxels.  The
same procedure is used recursively.  The main advantage of this
procedure is that the volume of images is reduced by a factor of
8 at each coarse-beading stage.  Therefore, the first iterations
of the procedure have a negligible cost as compared to the final
step where the full images are used.

\subsection{Differences with respect to 2D implementation}
\label{2D-3D}

The specificity of three-dimensional images is the large amount of
data used in the analysis, and hence the procedure is demanding in
terms of memory requirements and computation time. To reduce this
over-cost, variations as compared to two-dimensional
implementations~\cite{1847} have been considered:
\begin{itemize}
\item Since one generally deals with small strains between
consecutive images, one assimilates the reference and deformed
coordinate systems, as classically performed in linear elasticity.
Therefore, instead of computing the gradient of the original
image, a more precise estimate is obtained from the gradient of
the average between reference and (corrected) deformed image.
However, this amounts to computing all gradients as often as the
deformed image is corrected, and from these quantities, estimating
the matrix $M$ and second member $b$, is a rather costly
operation. This procedure was not implemented in three dimensions.
Rather, the above formulation is implemented directly, and since
the reference image $f$ remains invariant, the matrix $M$ is
computed once for all at the start of the procedure, and only the
second member is corrected.
\item The correction process of the deformed image to be performed
at each iteration involves a sub-pixel interpolation of the gray
levels. In two dimensions, a Fourier transform was used in order
to provide a simple interpolation of the data, convenient for
translation of zones.  However, to limit edge artifacts, each zone
was initially extended by a frame in which aliasing effects were
taken care of through a smooth connection between opposite edges.
In the present case,  this procedure would be too time consuming,
and thus a simple and fast (tri-)linear interpolation of gray
levels of the elementary voxel volume containing the required
coordinate is implemented.
\item The computation of the gradients is performed in
two-dimensions through the Fourier interpolation procedure alluded
to above. This is important to have consistent gradient and
interpolation procedures.  In three dimensions, since a linear
interpolation is used, a simple finite difference scheme is used.
\end{itemize}

\section{Application to foam tomography}
\label{se-perfo}

\begin{figure}
\begin{center}
 a){\epsfxsize0.45\textwidth \epsffile{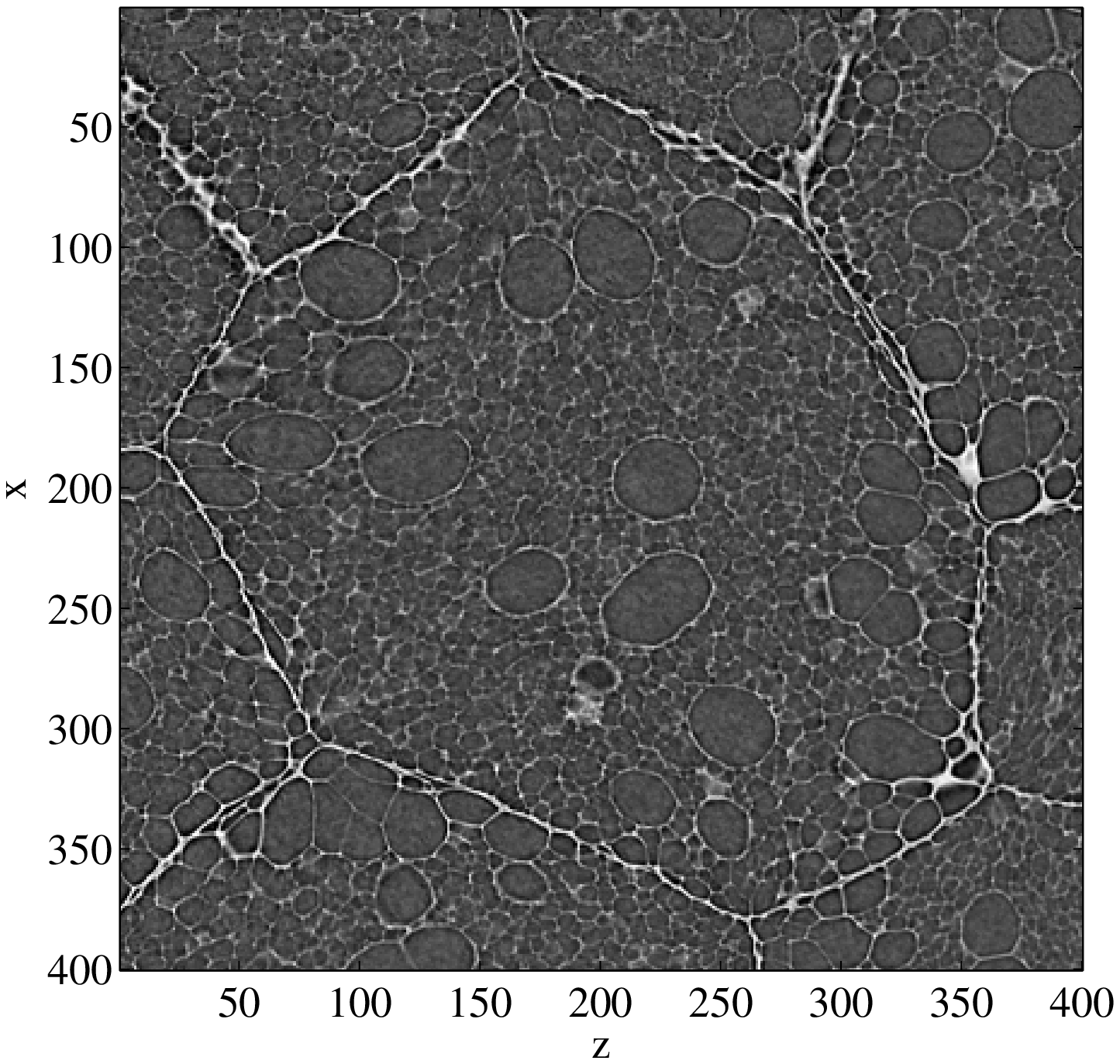}}
 \hfill
 b){\epsfxsize0.45\textwidth \epsffile{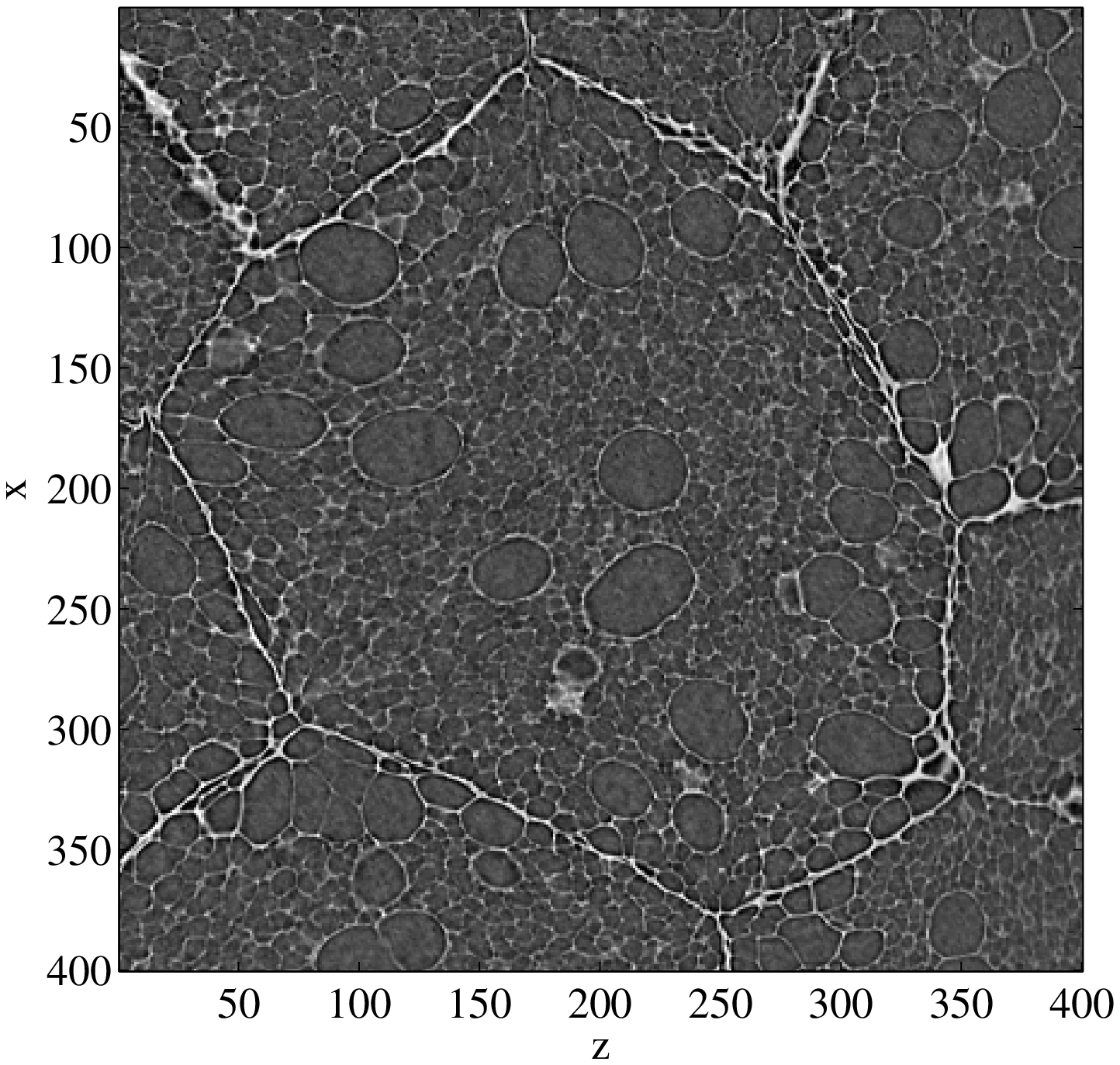}}
 c){\epsfxsize0.45\textwidth \epsffile{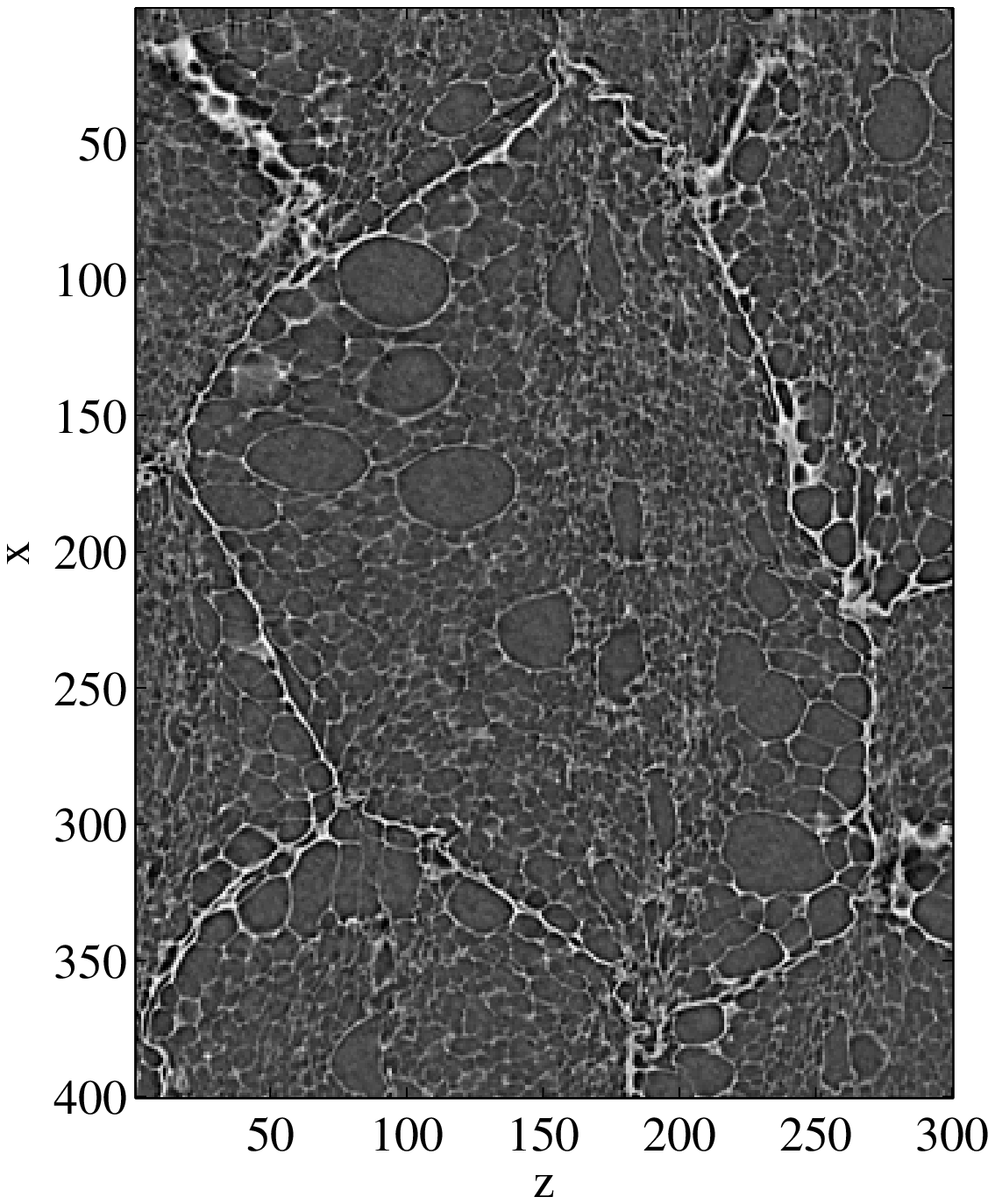}}
\end{center}
  \caption{Cut through the bead ``61'' in (a) the original state,
  (b) the $-10$\% compression state, and (c) the $-30$\% compression
  state. A strain localization is clearly perceptible in image (c),
  and can be also distinguished in image (b).}
  \label{fig:cuts}
\end{figure}

The previously described algorithm is now applied to the three
dimensional pictures obtained from XCMT of one bead extracted
from a complete specimen.  Three states were considered in the
experiments. The first one is the reference picture (before any
loading). The second is obtained from the sample macroscopically
strained at $-10$\% axial compression along the $z$-axis (just
after the first impact). The last one was compressed at $-30$\%
mean compression (after the second impact). The strain field of
the foam is particularly heterogeneous during the first stage of
the dynamic loading. In the companion paper~\cite{1974}, the
integral volumetric strain, tr$(\epsilon)$, at the bead scale was
estimated based on a bead wall location technique. Considering
only the bead extracted from the sample (labelled ``61'', for
more details, see Ref.~\cite{1974}) the  average volume change of
this bead is respectively $-2.7$\% and $-17$\% for the two
studied deformation states. Figure~\ref{fig:cuts} shows three
cuts through the same bead studied in this article at the three
stages of deformation (macroscopically 0, $-10$ and $-30$\%
strain). One clearly sees in the last stage that the strain is
such that some cells collapsed. Moreover, one observes that the
strain field is very heterogeneous, and a (vertical) compression
band may be perceived. At a much more preliminary stage, the same
band can already be seen as initiating at the $-10$\% strain
stage. Unfortunately, the last pair of images was too much
deformed to allow for a satisfactory determination of the
displacement. In fact, a good matching of only part of the sample
was obtained, and the present localized compression band forbids
a global convergence. Thus this last sample will be left aside
from the present analysis, and we will only report on the first
image pair.

\subsection{Image characteristics}

Each tomographic image was initially acquired with a resolution
of $2048 \times 2048 \times 2200$~voxels whose physical size is
4.91~$\mu$m, encoded as 8-bit deep gray levels.  This represents
a considerable amount of information (\ie 9.2 Go), which is
extremely demanding in terms of memory (storage, RAM and graphic
memories). Therefore the entire specimen was not used but rather
a specific subset. We focused on a single bead (labelled ``61'').
The volume containing this bead was already large (\ie
$800^3$~voxels). A first data compression was performed reducing
the system size by a factor of 4 in each direction. No
re-projection of the mean density onto an 8-bit range was
performed; the direct sum of the gray levels of the $4^3$ block
of voxels was assigned to the equivalent macro-voxel. The dynamic
range corresponds to $14$-bit digitization. The actual dynamic
range is of the order of $16,000$, \ie quite close to the maximum
value ($16,384$). The physical size of those super-voxels is thus
four times larger than the original voxel, or 19.7 $\mu$m.

Last, in those images, a region of interest (ROI) is defined in
which the displacement field will be analyzed.  In the following,
a \revision{cubic} ROI of the order of \revision{$100$~voxels
wide in each direction} was defined, and centered in the
compressed image. The size of the ROI is adjusted to correspond
to an integer number of ZOIs (hence typically 104 or 96~voxels).
For such sizes, one could deal with ZOIs as small as $5$~voxels
on a laptop computer (Intel dual core Centrino processor, 2 Go
RAM) without encountering memory limitations. For larger ZOIs,
\ie 16-voxel ZOIs, one could reach larger domain sizes (\eg
\revision{$176$ voxels in each direction}). Computation time
turned out not to be a limitation factor. A maximum CPU time of
the order of 10~min (for ROI sizes of 100~voxels, and a ZOI size
of 8~voxels) was sufficient to reach convergence.  For a ROI size
of 176~voxels, and ZOI size of 16~voxels, the computation time
reached about 1 hour.

\subsection{A priori performance}

Before studying the displacement field between two images,  it is
important to evaluate the level of uncertainty attached to both
the natural texture of the image and the algorithm used. Integer
valued displacements do not involve any approximation  in the
determination of the gray level value.  The only difficulty is to
be able to capture the displacement without being trapped in
secondary minima of the objective function.  The most stringent
limitation comes from gray value interpolation at non-integer
positions.  Therefore, in order to assess the uncertainty
resulting from this interpolation guess, a uniform displacement
of $(0.5,0.5,0.5)$ voxel is prescribed on the reference image to
produce an artificial ``deformed'' image.  \revision{This
subvoxel displacement is performed in Fourier space by a phase
shift operation.} The procedure is then run blindly on this pair
of images. The systematic error is measured from the spatial
average of the determined displacement. The uncertainty is
measured from the standard deviation of the displacement field
about its mean value. It turns out that the systematic error is
quite small as compared to the uncertainty. Only the latter is
reported herein.

\begin{figure}
\centerline{\epsfxsize0.65\textwidth \epsffile{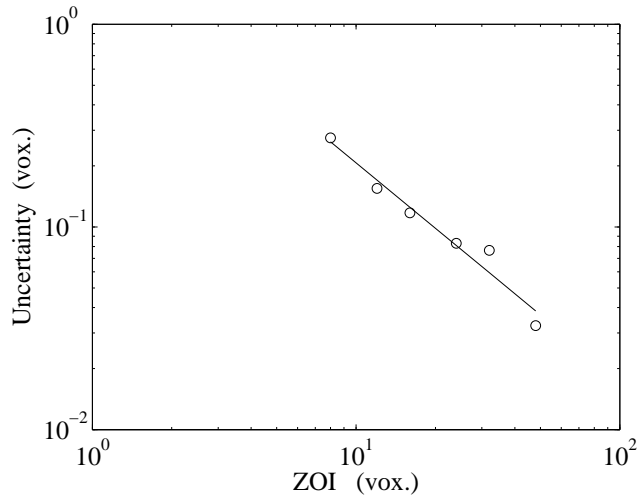} }
  \caption{Log-log plot of the uncertainty in displacement for the artificial case
  of a uniform translation of 0.5 voxel in each spatial direction,
  versus ZOI size.  Data points are shown as ($\circ$) and the straight
  line is a power-law fit to the data.}
  \label{fig:uncertainty}
\end{figure}

The quantification of the uncertainty with respect to the ZOI size
is very important because it allows one to determine the optimal
ZOI size to be used for a given constraint on the error bars.
Large ZOIs will enable one to determine accurate displacements
because of the large number of voxels involved in each element.
Conversely, they are unable to capture complex displacement fields
with rapid spatial variations.  Conversely, a small ZOI will be
flexible enough to follow large displacement gradients, but will
give less accurate displacement evaluations.

The uncertainty (here defined by means of the standard deviation
$\sigma(U)$ is typically observed in two dimensions to vary as a
power-law of the ZOI size, $\ell$ for FFT-DIC~\cite{1632} or
Q4-DIC~\cite{1847}
  \be
  \sigma(U) \approx A^{\alpha+1} \ell^{-\alpha}
  \ee
with an exponent $\alpha$ of the order of 1 for FFT-DIC and of
order 2 for Q4-DIC and very good textures.

The average uncertainty on the displacement may be estimated to
be of order half of $\sigma(U)$ as computed for a displacement of
0.5 voxel in each direction~\cite{1632,1847}.
Figure~\ref{fig:uncertainty} shows that a reasonable power-law is
observed in three dimensions, where a best fit through the data
produces the plain line on the graph, with $A=2.4$~voxels and
$\alpha=1.07$.   The prefactor of the mean uncertainty is thus
$A/2$, which compares quite well with the corresponding value in
two dimensions, where the prefactor is approximately 1~pixel (or
$A \approx 2$~voxels). The exponent, $\alpha$, is a bit
disappointing as compared to two dimensional performances. The
origin of this small exponent is not yet clear. It may come from
the poor approximation used in interpolation and gradients, or be
the result of the specific texture of the image. When dealing
with more pictures, it will be possible to answer that question.

In order to quantify the strain uncertainty (last column of
Table~\ref{tab:results}), the mean strain is estimated over the
entire region of interest for the rigid body translation of 0.5
pixel in each direction. The uncertainty is defined as the {\em
maximum} of the absolute value of the principal strain. As
expected, the strain uncertainty decreases with the ZOI size.
However, it remains less than $7 \times 10^{-3}$ for all used ZOI
sizes, \ie small enough to allow us to address the case under
study.

\section{Results for one loading step}
\label{se-results}

\subsection{Matching gap}

In order to evaluate quantitatively the difference between two
images, a matching gap is defined as
  \be
  R=\frac{\sigma(f-g)}{\max(f)-\min(f)}
  \ee
where $\sigma$ denotes the root mean square value of the picture
difference. It is a dimensionless indicator of the distance
between two images, reaching 0 only for identical images. The
initial gap between the two images over the region of interest is
9.49~\%. The first step, which consists in estimating the mean
rigid body translation, reduced the gap to 8.65~\%. Good quality
optical images are matched up to a final gap of the order of
1~\%. In our case, the gray levels should be comparable since,
for all images, the same affine transformation to 8-bit
conversion is performed after the reconstruction has been
performed. Slight difference may however occur due to
interferences, (a coherent X-ray beam was used at ESRF) but this
may give rise to about 2 gray level variations (or about 1\%) in
the original images.

In the present case, at convergence, the gap drops to about 5~\%
between the reference and the corrected image.
Table~\ref{tab:results} summarizes the final gap for different ZOI
sizes.  As expected, as the ZOI decreases in size, more and more
degrees of freedom are available to reduce the gap. However, the
level reached for the smallest ZOI sizes seems to indicate that
the incompressible level of mismatch between both images is
reached. Although somewhat larger than for two-dimension optical
images, the gain with respect to the initial gap is significant
enough to indicate that the matching presumably cannot be improved
further with the present algorithm.

\begin{table}
\begin{center}
\caption{Influence of the ZOI size on various quantities of
interest: mean displacement, matching gap, volume change, axial
strain in the compression direction and strain
uncertainty.}\label{tab:results}
\bigskip
\begin{tabular}{|c|c|c|c|c|c|c|c|}
\hline ZOI size & $\langle U_x\rangle$ & $\langle U_y\rangle$ &
$\langle U_z\rangle$ &
Gap &$\quad$tr$(\varepsilon)\quad$& $\quad\varepsilon_{zz}\quad$
& $\delta\epsilon$\\[-15pt]
(vox.) & (vox.) & (vox.) & (vox.) & (\%) & (--) & (--)& (--)\\
\hline\hline
16 & $-2.28$ & $-0.77$ & $-2.99$ & 5.51 & $-0.028$ & $-0.038$ & $0.003$\\
\hline
12 & $-2.27$ & $-0.77$ & $-3.00$ & 5.16 & $-0.028$ & $-0.037$ & $0.004$\\
\hline
8  & $-2.27$ & $-0.77$ & $-2.99$ & 4.88 & $-0.025$ & $-0.034$ & $0.005$ \\
\hline
6  & $-2.27$ & $-0.78$ & $-2.99$ & 4.76 & $-0.022$ & $-0.032$ & $0.007$ \\
\hline
\end{tabular}
\end{center}
\bigskip
\end{table}


\subsection{ZOI size sensitivity}

To check for the stability of the results with respect to the ZOI
size, different indicators extracted from the displacement fields
are proposed in Table~\ref{tab:results}. First, the mean
displacement estimated from the arithmetic average $\langle \ldots
\rangle$ of the nodal displacements, along the three spatial
directions, are indicated for ZOI sizes ranging from 16 to
6~voxels. The displacement averages are stable and only fluctuate
by about 0.01 voxel (corresponding to 0.197 $\mu$m).

To further probe the stability of the displacement results, the
mean strain is computed over the entire region of interest.
\revision{The mean strain is classically defined as the volume
(arithmetic) average of the strain.  In the chosen
discretization, the strain is constant in each element. So that
the mean strain is also the average of the strain over elements.
} In this case, a simple check shows that the inner nodes do not
contribute to this estimate and hence only the boundary nodal
values matter. However, because these nodes are located on the
boundary, they are much less constrained than inner nodes, and
hence much susceptible to noise.  In order to reduce this
spurious effect, the strain estimates are performed by ignoring
the outermost elements.  One consequence of this however is the
fact that the domain over which the strain is estimated is not
quite identical when the ZOI size changes.  It is observed that
the mean volume change, \ie div($\mathbf U$), or
tr($\varepsilon$), which is a compression of order $-2.5$\% has a
tendency to decrease with the ZOI size. However this decrease is
only $6 \times 10^{-3}$ and this estimate of the strain is in
agreement with the result obtained by measuring the volume change
of iso-surfaces representing the bead 61 ($-2.7$\%, see
Ref~.\cite{1974}). We will see in Section~\ref{subsec:ZOI size}
that this trend has a simple interpretation. One observes also
that the axial strain along the direction of compression follows
exactly the same tendency, and hence the transverse area change,
$\varepsilon_{xx}+\varepsilon_{yy}$, is approximately constant
(fluctuation of the order of $10^{-3}$).

\subsection{Check of correction from mid-plane cuts}

\begin{figure}
\centerline{a)\epsfxsize0.45\textwidth
\epsffile{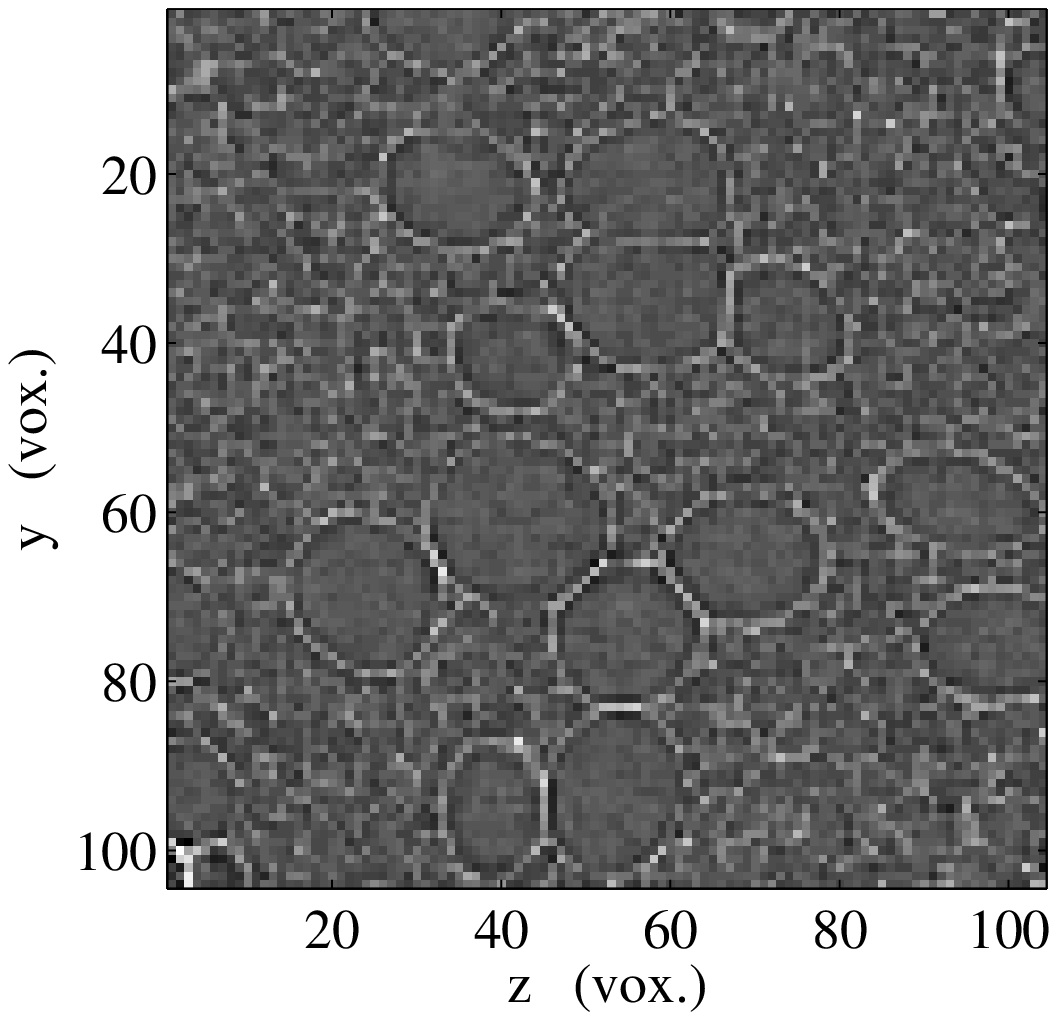}\hfill b)\epsfxsize0.45\textwidth
\epsffile{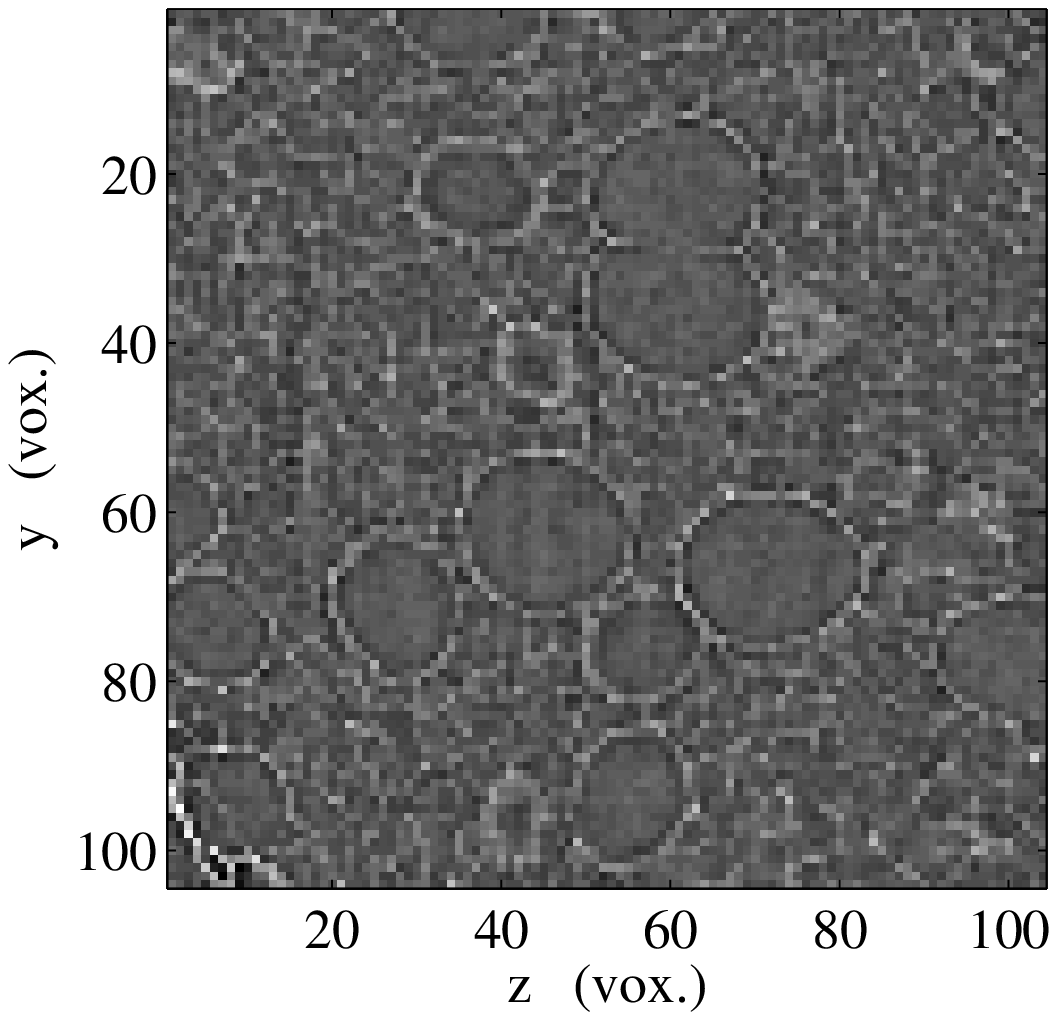}} \centerline{c)\epsfxsize0.45\textwidth
\epsffile{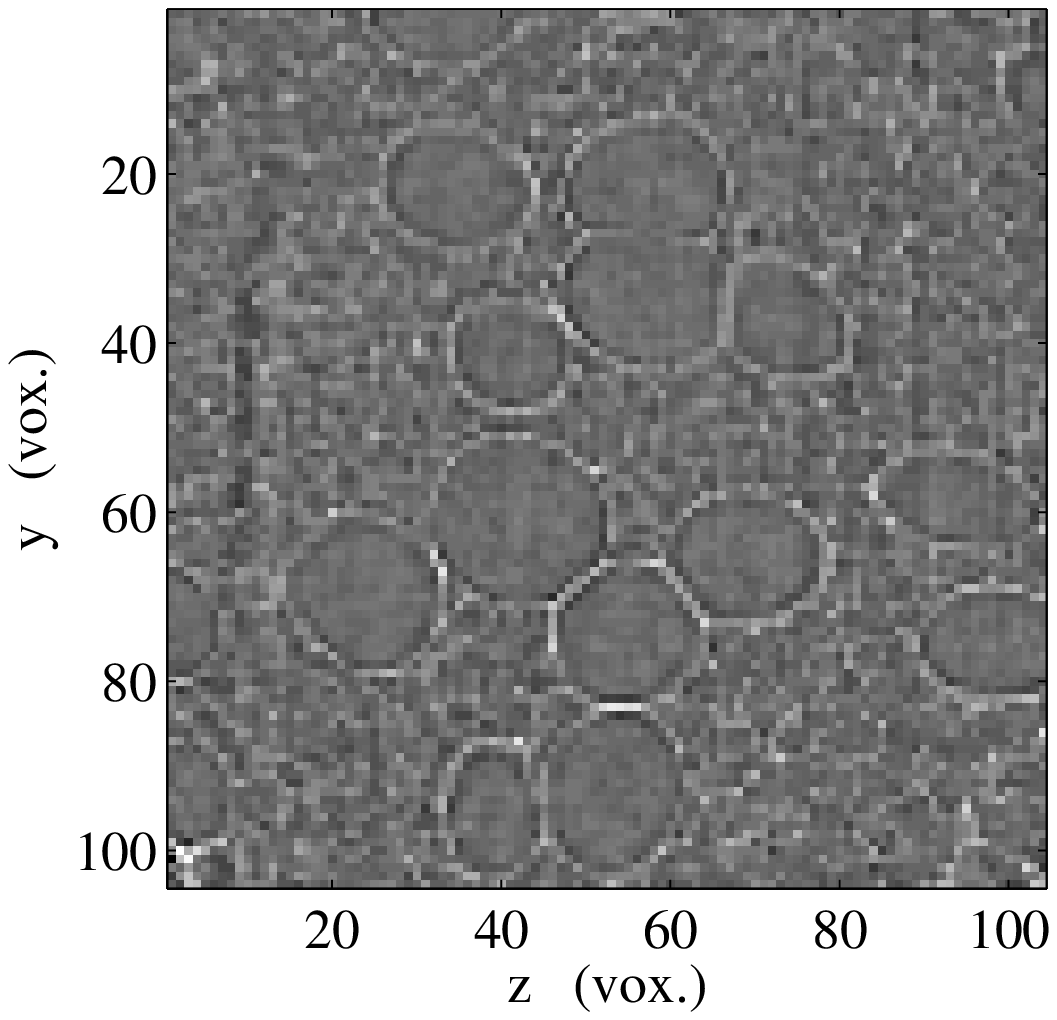}}
  \caption{Reference (a), Deformed (b) and Corrected (c) cut
  through the specimen for $x=100$~voxels.}
  \label{fig:Cut_x100}
\end{figure}

\begin{figure}
\centerline{a)\epsfxsize0.45\textwidth
\epsffile{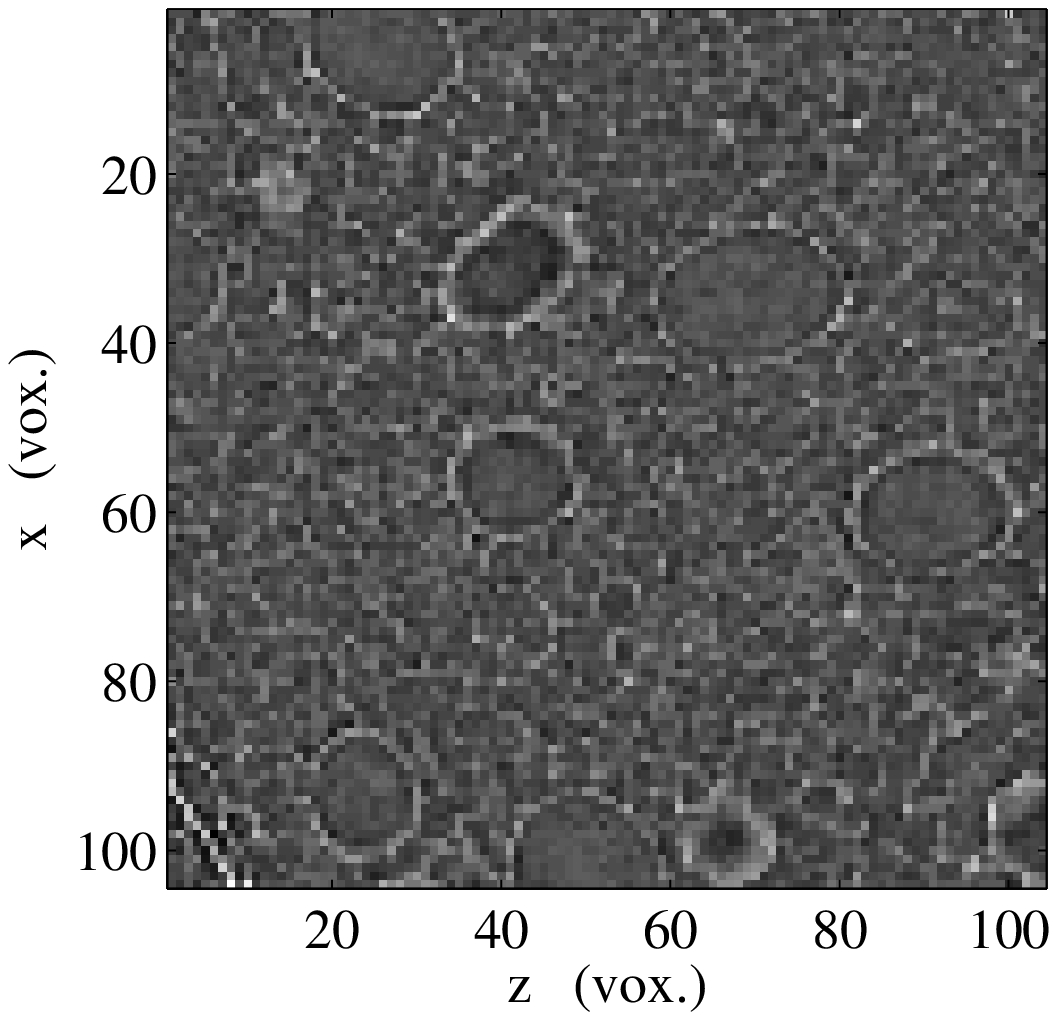}\hfill b)\epsfxsize0.45\textwidth
\epsffile{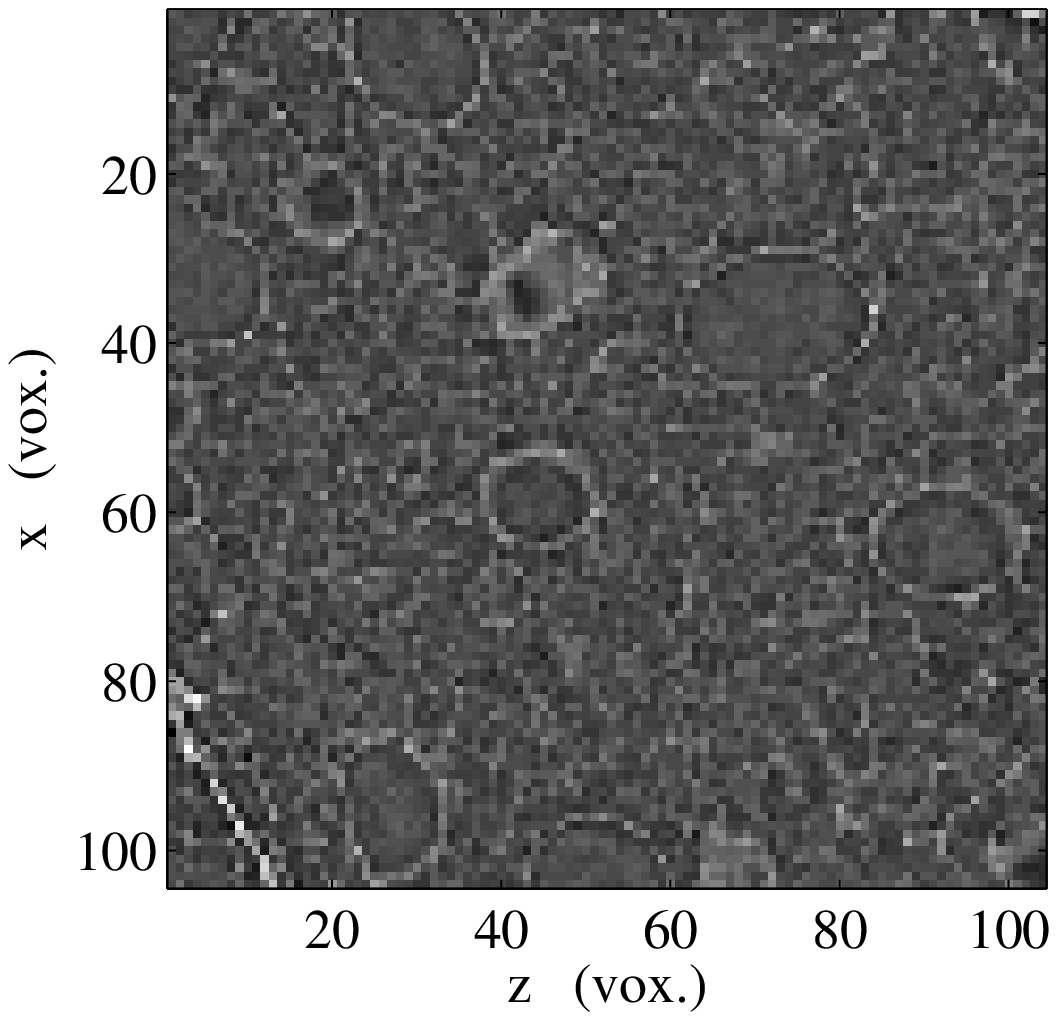}} \centerline{c)\epsfxsize0.45\textwidth
\epsffile{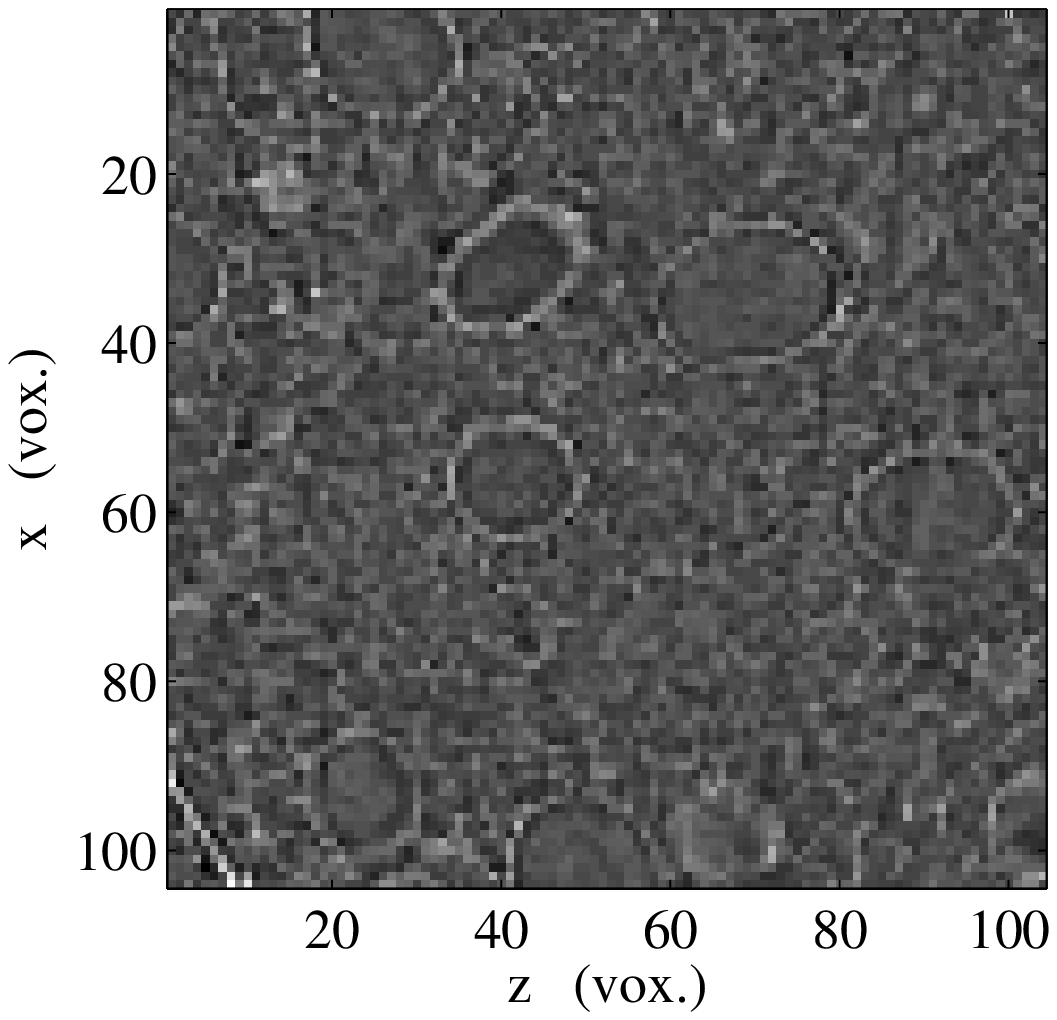}}
  \caption{Reference (a), Deformed (b) and Corrected (c) cut
  through the specimen for $y=100$~voxels.}
  \label{fig:Cut_y100}
\end{figure}

\begin{figure}[h!]
\centerline{a)\epsfxsize0.45\textwidth
\epsffile{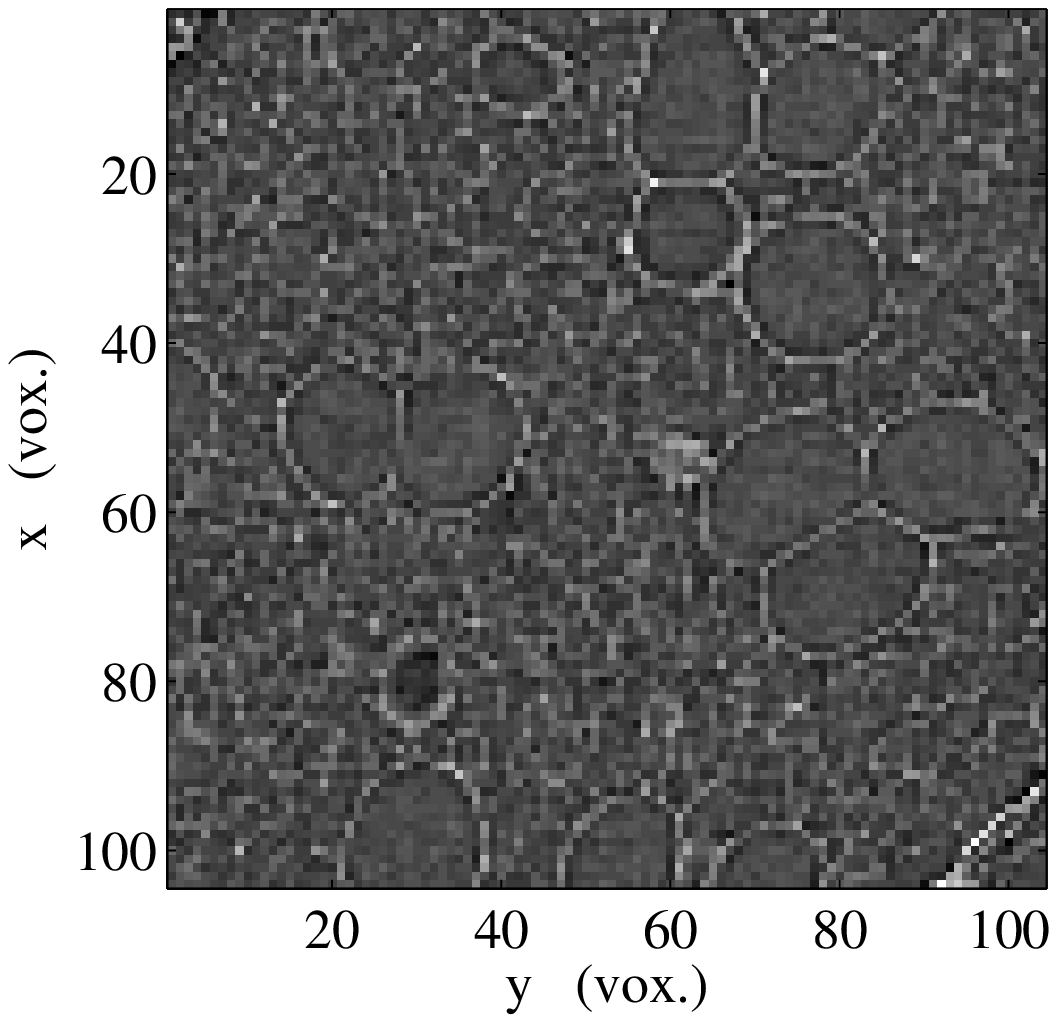}\hfill b)\epsfxsize0.45\textwidth
\epsffile{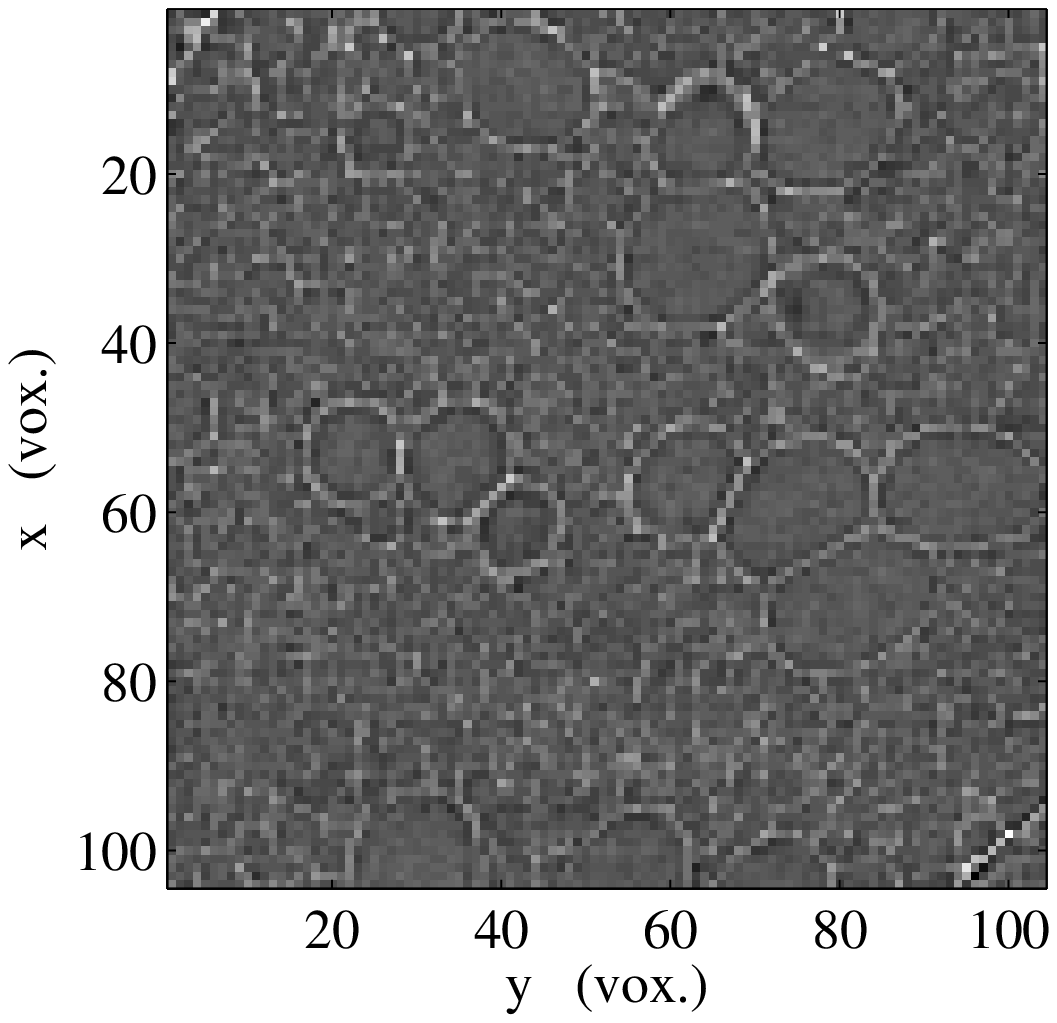}} \centerline{c)\epsfxsize0.45\textwidth
\epsffile{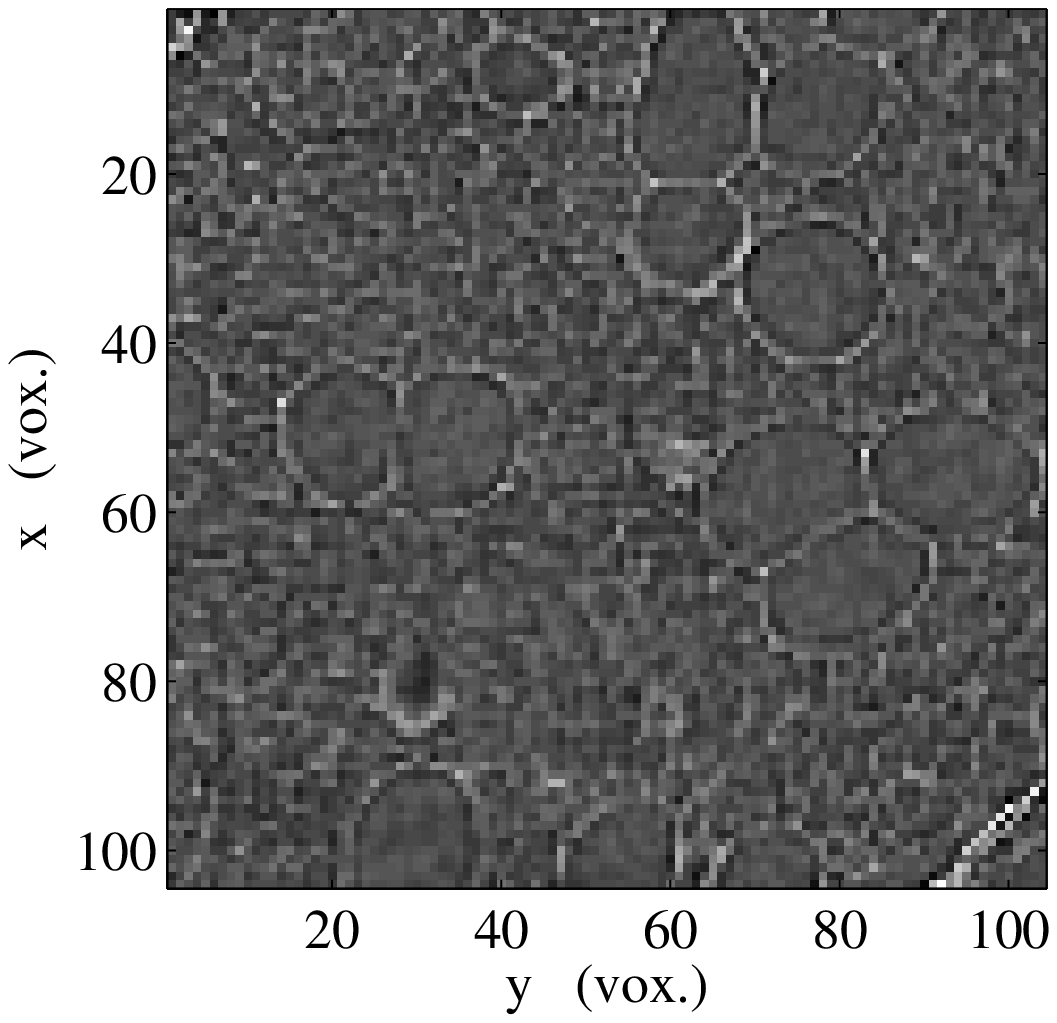}}
  \caption{ Reference (a), Deformed (b) and Corrected (c) cut
  through the specimen for $z=100$~voxels.}
  \label{fig:Cut_z100}
\end{figure}

Although it is important to deal with a single number such as the
matching gap to evaluate the quality of the displacement
determination, a global quantity may hide different types of
deviates, from a uniformly distributed difference, to a large but
exceptional difference. To have a more faithful appreciation of
the quality of the determination, Figures~\ref{fig:Cut_x100},
\ref{fig:Cut_y100} and \ref{fig:Cut_z100} show three midplane cuts
(normal to $x$, $y$ or $z$) of the reference image (a), as
compared to the as-received deformed image  (b), and the corrected
deformed image (c) once the determined displacement has been taken
into account. Had the determination and image acquisition been
perfect, then the section views (a) and (c) would have been
identical. Although very significant differences are perceived on
the images (a) and (b), the similarity between (a) and (c) is
clear.

To further check this similarity, a two-dimensional image
correlation analysis (Q4-DIC) was run on the three (a)-(c) pairs.
This software is similar in spirit to the one used in three
dimensions, and for the image pairs (a)-(c) ideally the
displacement field should be zero.  It is worth noting that both
codes are totally independent, and use some slightly different
prescriptions for the interpolation and gradient computation (as
discussed in Section~\ref{2D-3D}).  This may be responsible for a
non-zero answer. The same ZOI size was used as for the three
dimensional computation (\ie 8 pixels). The obtained displacement
field was less than 0.02 pixel for all three cuts, and with no
visible spatial correlation, (the displacement field seemed
similar to white noise).  This check is thus considered as an
additional validation of the result.

Let us also note that in Figure~\ref{fig:Cut_x100} the deformed
image as well as the corrected one display marked vertical lines
that seem to be artifacts of the reconstruction.  This illustrates
the previous discussion on relatively high level of matching gap
between images at convergence.

\subsection{Influence of ZOI size}
\label{subsec:ZOI size}

\begin{figure}
\centerline{a) \epsfxsize0.45\textwidth
\epsffile{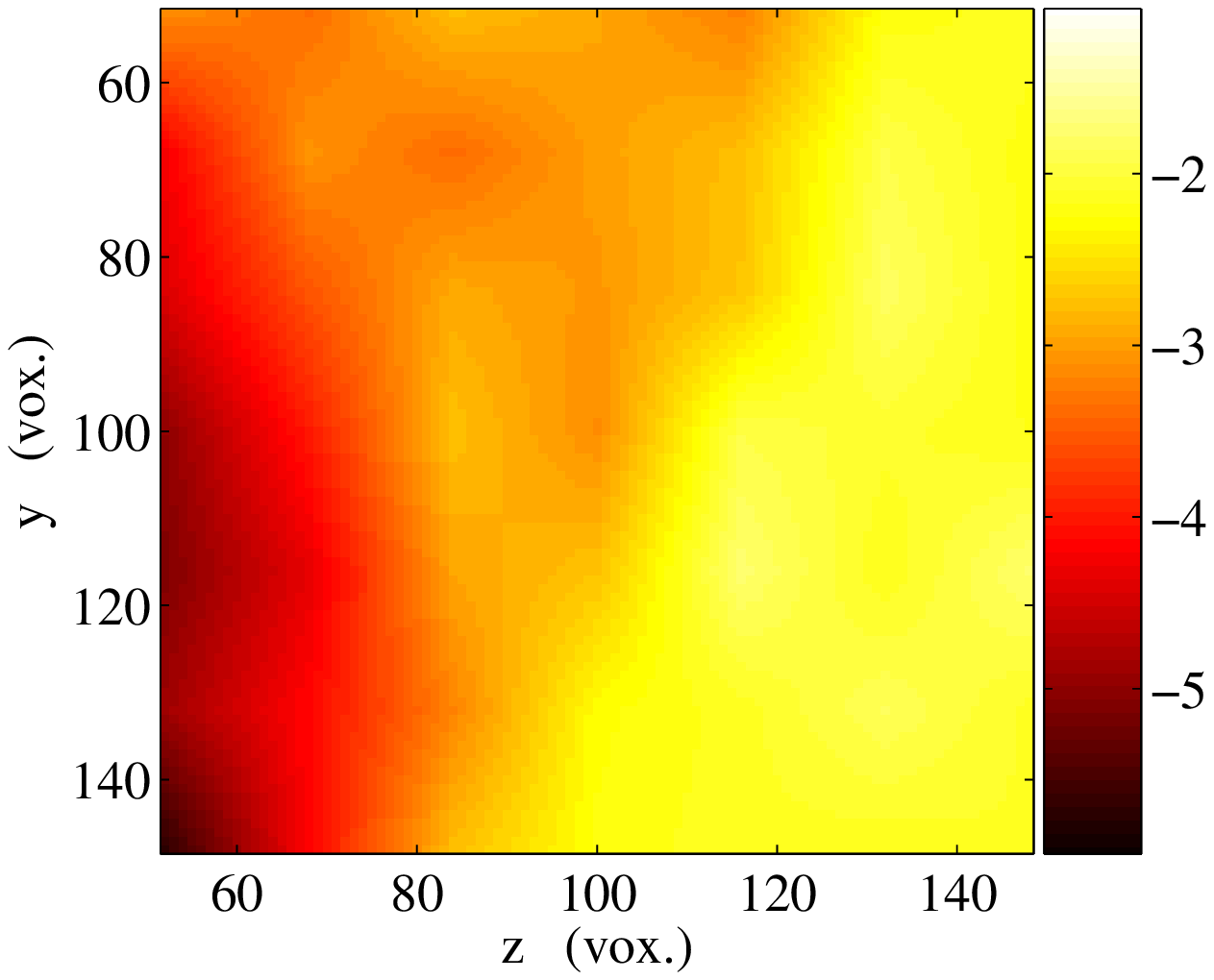}\hfill \epsfxsize0.45\textwidth
b) \epsffile{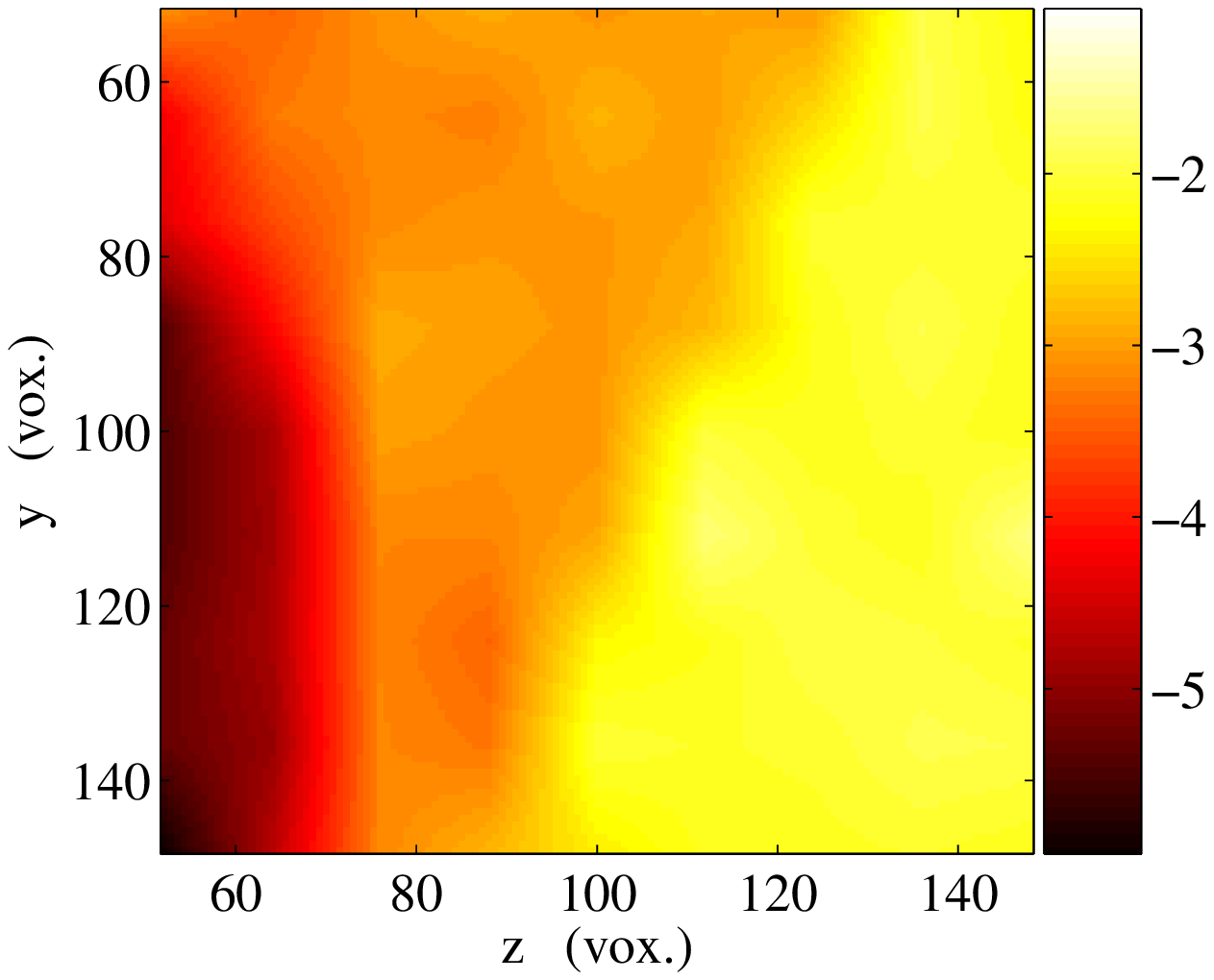}}
\centerline{\epsfxsize0.45\textwidth c)
\epsffile{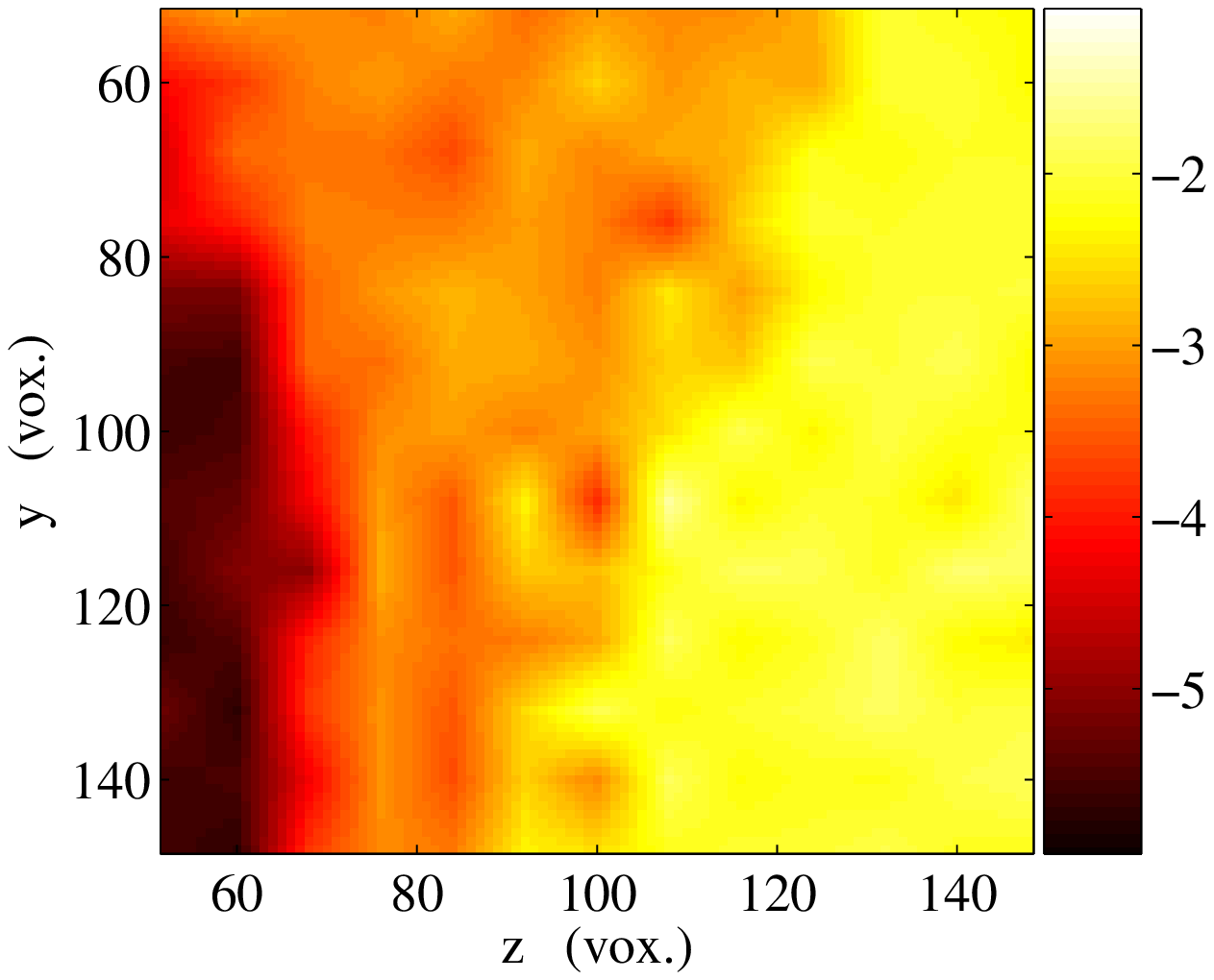}\hfill \epsfxsize0.45\textwidth
d) \epsffile{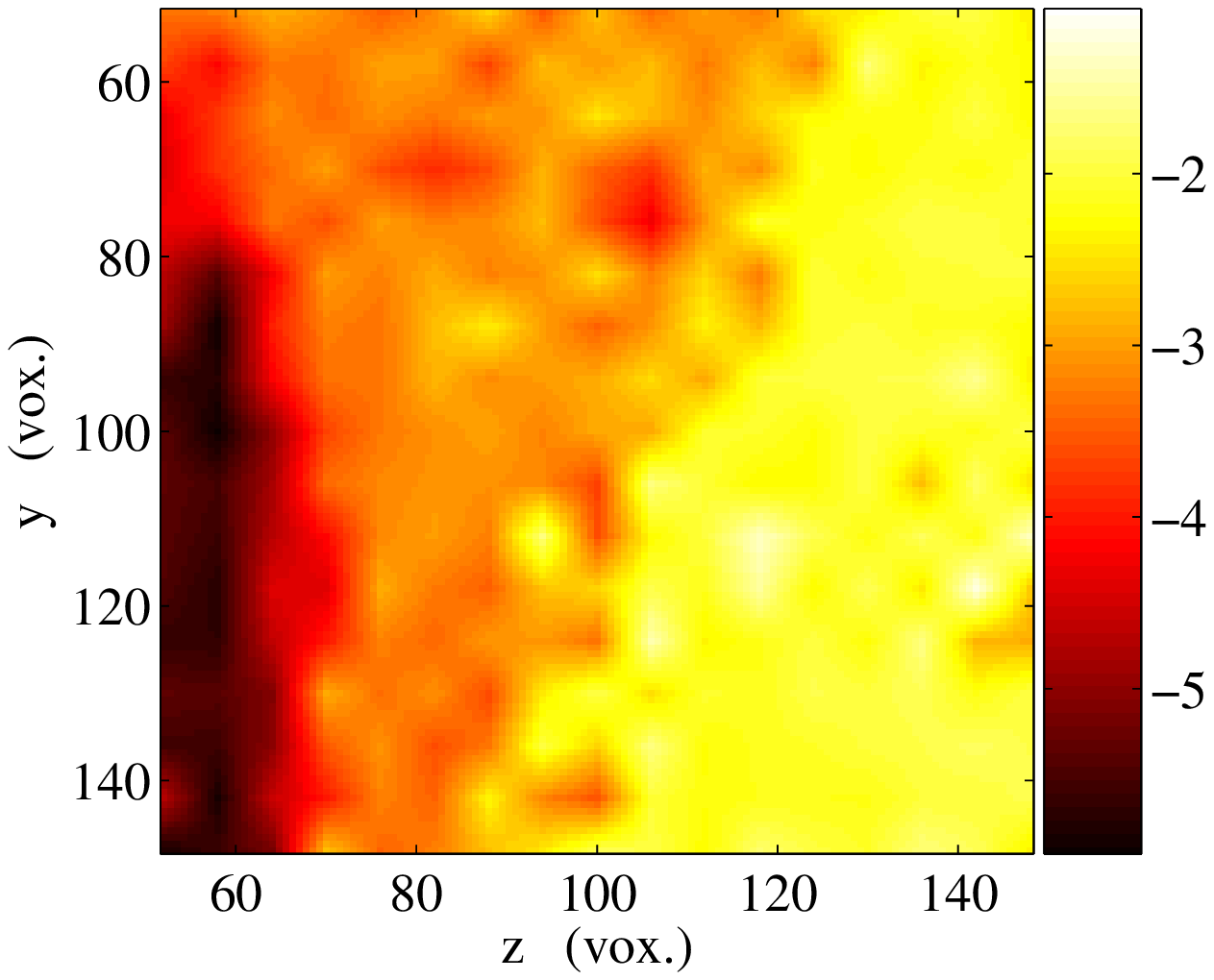}}
  \caption{Comparison between the $U_z$-component of the displacement
  field in the $x=100$~voxels plane, determined using different ZOI sizes
  a) 16; b) 12; c) 8; d) 6~voxels.  The displacement scale is in voxel
  (1 vox.~$=19.7 \mu$m).}
  \label{fig:Uz_x100}
\end{figure}

One aspect associated with three dimensional analyses of
displacements is the fact that it is difficult to visualize three
dimensional vector fields. Hence we do not show such graphs.
However, to give some indications of the measured displacement
field, 2D maps of the component of the displacement field that
has the largest mean gradient, namely $U_z$, along the
compression axis is shown. A reference plane corresponding to the
mid-plane normal to the $x$-axis ($x=100$~voxels) is chosen.
Figure~\ref{fig:Uz_x100} shows the corresponding maps of
$U_z(x=100~\mathrm{voxels},y,z)$ for four values of the ZOI size,
ranging from 16 to 6 voxels in each direction.  It is to be noted
that the large scale features of the displacement are preserved.
\revision{It is also observed that small ZOIs are required to
capture the sharp gradients, but they unfortunately enhance the
noise level in the displacement measurements.}

\begin{figure}
\centerline{a) \epsfxsize0.45\textwidth
\epsffile{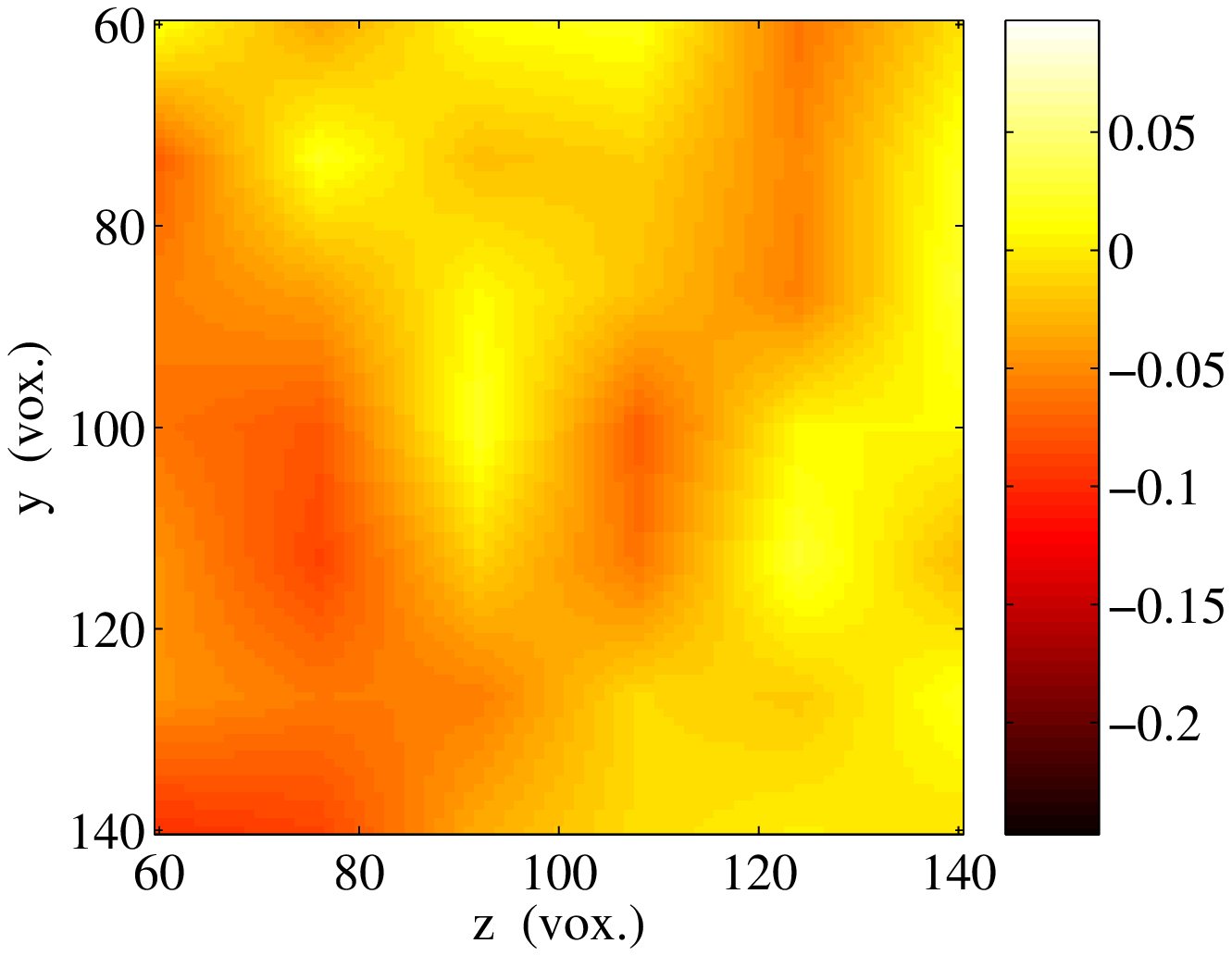}\hfill \epsfxsize0.45\textwidth b)
\epsffile{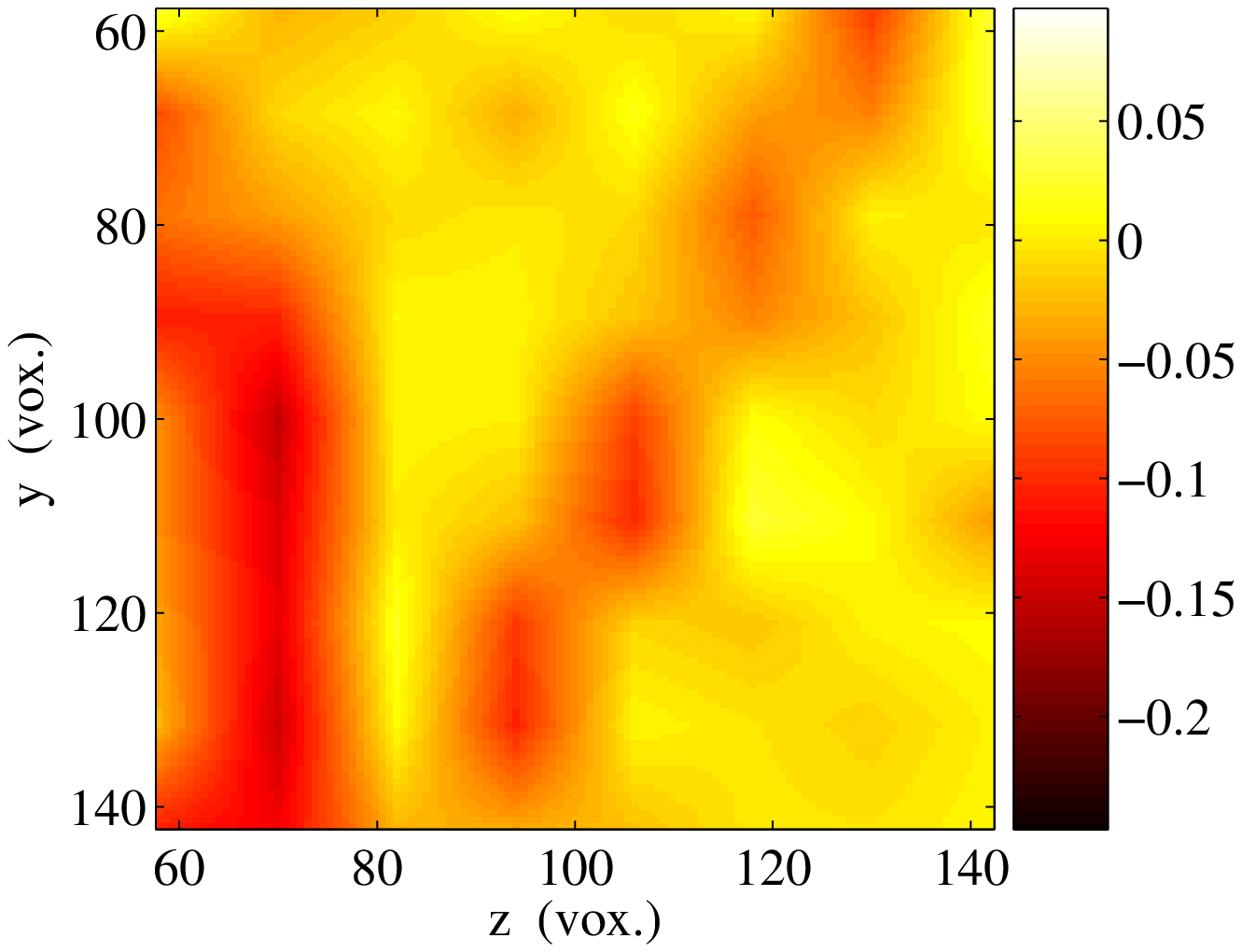}} \centerline{\epsfxsize0.45\textwidth
c) \epsffile{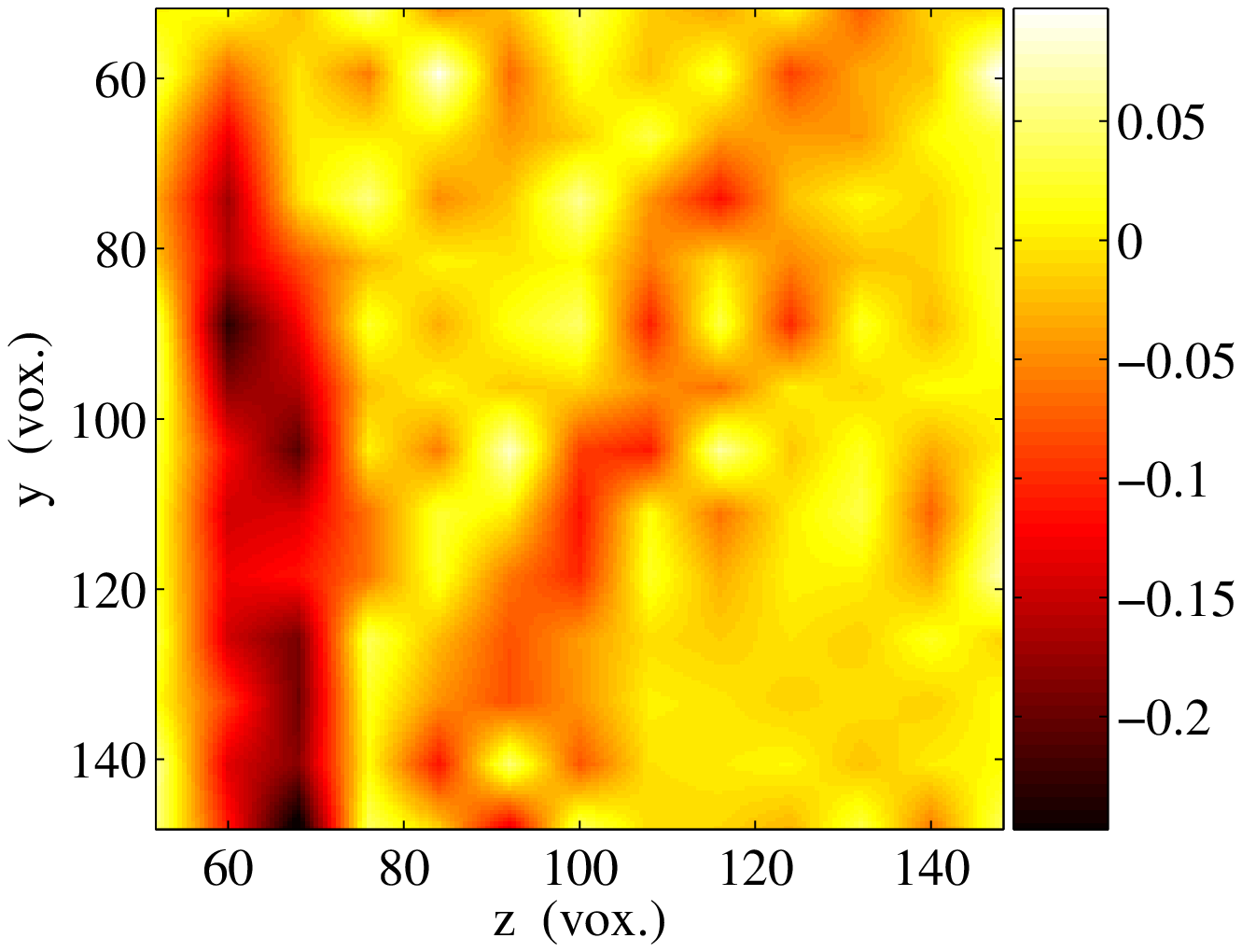}\hfill \epsfxsize0.45\textwidth d)
\epsffile{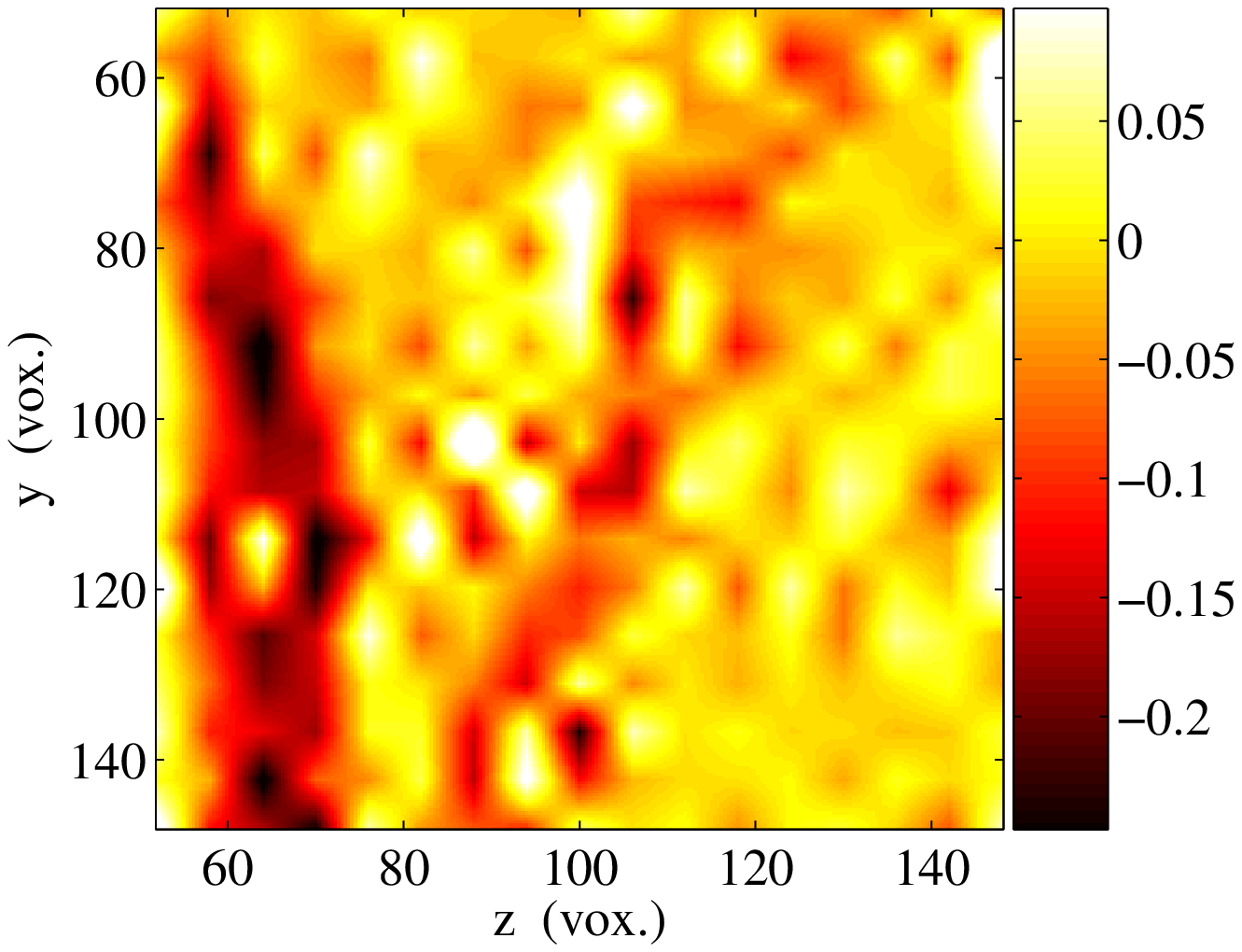}}
  \caption{Comparison between the $\varepsilon_{zz}$-component of the
strain field in the $x=100$~voxels plane, determined using
different ZOI sizes a) 16; b) 12; c) 8; d) 6~voxels.  In order to
decrease the visual impact of the discretization, the maps have
been linearly interpolated between element centers where the
element strain has been imposed.}
  \label{fig:epszz_x100}
\end{figure}

\revision{In order to better illustrate this effect, the
corresponding strain maps of the $\varepsilon_{zz}$ component are
shown in Figure~\ref{fig:epszz_x100}.  The chosen discretization,
C8P1 elements, implies that the strain should be a constant in
each element.  However, such maps are difficult to compare since
the visual impression is limited by the inter-element
discontinuities.  It was chosen to use a linear interpolation
between strains assigned to the center of each element.  Let us
however note this is a simple rendering effect which does not
quite corresponds to the discretization.  Note the very large
amplitude of the strains distribution shown in this figure, and
the consistency between different ZOI sizes. It is obvious that
the large ZOI of 16 voxels leads to a diffusion of the strain
over large sizes and hence cannot estimate faithfully sudden
variations. For sizes of 12 or 8 voxels, two features become well
pronounced, namely a compression band located on the left of the
domain with strains reaching -25\%, and an inclined shear band.
For a 6-voxel ZOI, the noise level becomes prohibitive to rely on
the quantitative estimates of the strains, but the same features
can be seen.}

It was previously mentioned that such \revision{compression or
shear bands} could be perceived (qualitatively) between the
reference (a) and $-10$\% deformed (b) specimen, and was much
more marked for the $-30$\% deformed sample
(Figure~\ref{fig:cuts}). Cells located on the left side of
Fig.~(\ref{fig:cuts}c) and along a median vertical cut have
collapsed. The phenomenon of bead wall buckling can also be seen
clearly on these last pictures. It is worth emphasizing that such
an abrupt displacement gradient is a very difficult test case
where a number of classical two-dimensional image correlation
code would fail to detect.

The fact that the strains are localized has a direct impact on
the evaluation of their mean values.  It was mentioned above that
the effective size over which the mean strain is computed depends
on the ZOI size, because all elements adjacent to the boundary
were discarded.  If the strain field were homogeneous, this would
have no consequence.  In the present case however, the opposite
assumption that the strain be negligible over the removed element
leads to a systematic correction of the order $(1-2\ell/L)$ where
$L$ is the ROI size. \revision{A mean axial strain of order
$-2.8\%$ {\em independent} of the ZOI size, is obtained when this
first order correction is taken into account.}

The amplitude of the displacement is also to be noted, namely,
the level spanned by the axial displacement is about 10 voxels,
or 530$\mu$m, which is already quite large for DIC techniques
even if strains were not localized. Let us also observe that as
the ZOI size decreases, the displacement field appears to become
\revision{more noisy}. This is an illustration of the uncertainty
that increases as the ZOI size decreases, as shown in
Fig.~\ref{fig:uncertainty}.

\section{Conclusion}

A new approach was developed to determine 3D displacement fields
based on the comparison of two CT scans. The sought displacement
field is decomposed onto a basis of continuous functions using
C8P1-shape functions as proposed in classical finite element
methods. The latter corresponds to one of the simplest kinematic
descriptions. It therefore allows for a {\em complete}
compatibility of the kinematic hypotheses made during the
measurement stage and the subsequent identification / validation
stages, for instance, by using finite element
simulations~\cite{1985}. Since a linearized functional is
considered, a multi-scale algorithm is implemented to allow for
small as well as medium range displacements (\ie at least of the
order of the element size).

The performance of the algorithm was tested on a reference image
obtained by CT to evaluate the reliability of the estimation,
which is shown to allow for either a reasonable accuracy for
homogeneous displacement fields, or for a very well resolved
displacement field down to element sizes as small as 6~voxels.
The displacement field is analyzed on a foam sample where a
localization band develops in uniaxial compression. For element
sizes ranging from 16 down to 6~voxels, the displacement field is
shown to be reliably determined. As far as mean strains are
concerned, consistent results were obtained. Strain localization
could be observed when analyzing more locally the displacement
maps. The analysis is therefore operational and reliable to deal
with CT pictures. However, average strain levels as high as
$-30$\% could not be analyzed even with the multi-grid
implementation. Scans at intermediate strain levels should have
been performed.

The application of this algorithm to study the behavior of
cellular materials is encouraging.  It has been possible to
highlight and identify the heterogeneity of the strain field in a
foam. Previous works have shown the heterogeneity of the
deformation between beads, at the mesoscopic scale. With this new
approach, by combining 3D-DIC and XCMT, the heterogeneity of the
strain inside a bead has been shown at a scale close to that of
the cell scale. The information will be useful to improve foam
models and to better understand the response of this cellular
material used in sandwich structures. The phenomena observed and
quantified from this method have to be taken into account in a new
model. Although these results are very encouraging, in order to
follow the entire development of the strain field, it appears
necessary to have a more progressive series of image with a more
limited displacement difference between successive scans so as to
be compatible with the limits of the presented DIC algorithm.

\section*{Acknowledgements}
This work is part of a project (PHOTOFIT) funded by the Agence
Nationale de la Recherche. We acknowledge the European Synchrotron Radiation
Facility for provision of synchrotron radiation facilities and we would like
to thank Joanna Hoszowska for assistance in using beamline BM05.

\end{document}